\documentclass[sigconf,balance=false]{acmart}
\usepackage{popets}

\usepackage[figuresright]{rotating}
\usepackage{amsmath,amsfonts}
\usepackage{algorithmic}
\usepackage{graphicx}
\usepackage{textcomp}
\usepackage{xcolor}
\usepackage{forest}
\usepackage{rotating}
\usepackage{caption}
\usepackage{multirow}
\usepackage{graphics}
\usepackage{tabularx}
\usepackage[skins]{tcolorbox}
\usepackage{float}
\usepackage{tabularray}
\usepackage{tikz}
\usepackage{balance}
\usepackage{adjustbox}
\usepackage{color, colortbl}
\usepackage{tabu}
\usepackage{xcolor}
\usepackage{paralist}
\usepackage{pifont}
\usepackage{varwidth}
\usepackage{cancel,soul}
\newcommand{\cmark}{\ding{51}}%
\newcommand{\xmark}{\ding{55}}%
\usepackage{smartdiagram, wrapfig}
\usepackage{enumerate}
\usepackage[shortlabels]{enumitem}

\PassOptionsToPackage{colorlinks,pagebackref,breaklinks}{hyperref}

\usepackage[colorlinks,pagebackref,breaklinks]{hyperref}

\hypersetup{
	hypertexnames=true, linkcolor=blue, anchorcolor=black,
	citecolor=blue, urlcolor=blue  
}

\usetikzlibrary{positioning,shapes,shadows,arrows}
\usetikzlibrary{arrows.meta,angles,quotes}

\tikzstyle{roundrect} = [rectangle, rounded corners, minimum width=3cm, minimum height=1cm,text centered, draw=black]
\tikzstyle{roundrect2} = [rectangle, rounded corners, minimum width=1.5cm, minimum height=1cm,text centered, draw=black]
\tikzstyle{roundrect3} = [rectangle, rounded corners, minimum width=0.5cm, minimum height=0.5cm,text centered, draw=black]

\def\BibTeX{{\rm B\kern-.05em{\sc i\kern-.025em b}\kern-.08em
		T\kern-.1667em\lower.7ex\hbox{E}\kern-.125emX}}

\newcommand*\halfcircL[1][1ex]{%
	\begin{tikzpicture}
		\draw[fill] (0,0)-- (90:#1) arc (90:270:#1) -- cycle ;
		\draw[thick] (0,0) circle (#1);
\end{tikzpicture}}
\newcommand*\halfcircR[1][1ex]{%
	\begin{tikzpicture}
		\draw[fill] (0,0)-- (270:#1) arc (270:450:#1) -- cycle ;
		\draw (0,0) circle (#1);
\end{tikzpicture}}
\newcommand*\fullcirc[1][1ex]{\tikz\fill (0,0) circle (#1);} 

\usetikzlibrary{shapes.geometric}

\colorlet{linecol}{black!75}

\smartdiagramset{
	description text width = 5.4cm,
	description title width = 1.7cm,
	description title font = \scriptsize,
	description font = \footnotesize,
}

\tikzset{
	my rounded corners/.append style={rounded corners=2pt},
}

\definecolor{Gray}{gray}{0.9}
\definecolor{airforceblue}{rgb}{0.36, 0.54, 0.66}
\definecolor{aliceblue}{rgb}{0.94, 0.97, 1.0}
\definecolor{alizarin}{rgb}{0.82, 0.1, 0.26}
\definecolor{amber}{rgb}{1.0, 0.75, 0.0}
\definecolor{amber(sae/ece)}{rgb}{1.0, 0.49, 0.0}
\definecolor{applegreen}{rgb}{0.55, 0.71, 0.0}
\definecolor{babyblue}{rgb}{0.54, 0.81, 0.94}
\definecolor{ballblue}{rgb}{0.13, 0.67, 0.8}
\definecolor{beaublue}{rgb}{0.74, 0.83, 0.9}
\definecolor{bronze}{rgb}{0.8, 0.5, 0.2}
\definecolor{battleshipgrey}{rgb}{0.52, 0.52, 0.51}
\definecolor{bole}{rgb}{0.47, 0.27, 0.23}
\definecolor{bulgarianrose}{rgb}{0.28, 0.02, 0.03}
\definecolor{carolinablue}{rgb}{0.6, 0.73, 0.89}
\definecolor{ceil}{rgb}{0.57, 0.63, 0.81}
\definecolor{cerulean}{rgb}{0.0, 0.48, 0.65}
\definecolor{charcoal}{rgb}{0.21, 0.27, 0.31}
\definecolor{columbiablue}{rgb}{0.61, 0.87, 1.0}
\definecolor{coolblack}{rgb}{0.0, 0.18, 0.39}
\definecolor{darkcandyapplered}{rgb}{0.64, 0.0, 0.0}
\definecolor{darkbrown}{rgb}{0.4, 0.26, 0.13}
\definecolor{darkgray}{rgb}{0.66, 0.66, 0.66}
\definecolor{darkjunglegreen}{rgb}{0.1, 0.14, 0.13}
\definecolor{darktaupe}{rgb}{0.28, 0.24, 0.2}
\definecolor{frenchblue}{rgb}{0.0, 0.45, 0.73}
\definecolor{almond}{rgb}{0.94, 0.87, 0.8}
\definecolor{beige}{rgb}{0.96, 0.96, 0.86}
\definecolor{bisque}{rgb}{1.0, 0.89, 0.77}
\definecolor{black}{rgb}{0.0, 0.0, 0.0}
\definecolor{fluorescentorange}{rgb}{1.0, 0.75, 0.0}
\definecolor{ghostwhite}{rgb}{0.97, 0.97, 1.0}
\definecolor{antiquewhite}{rgb}{0.98, 0.92, 0.84}
\definecolor{anti-flashwhite}{rgb}{0.95, 0.95, 0.96}
\definecolor{snow}{rgb}{1.0, 0.98, 0.98}
\definecolor{whitesmoke}{rgb}{0.96, 0.96, 0.96}

\setcopyright{popets}
\copyrightyear{YYYY}

\acmYear{YYYY}
\acmVolume{YYYY}
\acmNumber{X}
\acmDOI{XXXXXXX.XXXXXXX}
\acmISBN{}
\acmConference{Proceedings on Privacy Enhancing Technologies}
\begin{document}

\title{Wildest Dreams:
 Reproducible Research in Privacy-preserving Neural Network Training}

\author{Tanveer Khan}
\affiliation{%
  \institution{Tampere University}
\city{Tampere}
\country{Finland}
}
\email{tanveer.khan@tuni.fi}

\author{Mindaugas Budzys}
\affiliation{%
  \institution{Tampere University}
\city{Tampere}
\country{Finland}
}
\email{mindaugas.budzys@tuni.fi}

\author{Khoa Nguyen}
\affiliation{%
  \institution{Tampere University}
\city{Tampere}
  \country{Finland}
}
\email{khoa.nguyen@tuni.fi}

\author{Antonis Michalas}
\affiliation{%
 \institution{Tampere University}
\city{Tampere}
\country{Finland}
}
\additionalaffiliation{
\institution{RISE Research Institute of Sweeden}
\country{Sweeden}
}
\email{antonios.michalas@tuni.fi}

\renewcommand{\shortauthors}{Khan et al.}

\begin{abstract}
Machine Learning (ML), 
addresses a multitude of complex issues 
in multiple disciplines, including social sciences, finance, 
and medical research. 
ML models require substantial computing power and are only as powerful as the data utilized. Due to the high computational cost of ML methods, data scientists frequently use Machine Learning-as-a-Service (MLaaS) to outsource computation to external servers. However, when working with private information, like financial data or health records, outsourcing the computation 
might result in privacy issues. Recent advances in Privacy-Preserving Techniques (PPTs) have enabled ML training and inference over protected data through the use of Privacy-Preserving Machine Learning (PPML). However, these techniques are still at 
a preliminary stage 
and their application in 
real-world situations is demanding. In order to comprehend the discrepancy between theoretical research suggestions and actual applications, 
 this work examines the past and present of PPML, 
 focusing on Homomorphic Encryption (HE) and Secure Multi-party Computation (SMPC) applied to ML. This work primarily focuses on the ML model's training phase, where maintaining user data privacy is of utmost importance. We provide a solid theoretical background that eases the understanding of current approaches and their limitations. We also provide some preliminaries of SMPC, HE, and ML. In addition, we present a systemization of knowledge of the most recent PPML frameworks for model training and provide a comprehensive comparison in terms of the unique properties and performances on standard benchmarks. Also, we reproduce the results for some of the surveyed papers and examine at what level existing works in the field provide support for open science. We believe our work serves as a valuable contribution by raising awareness about the current gap between theoretical advancements and real-world applications in PPML, specifically regarding open-source availability, reproducibility, and usability.
\end{abstract}


\maketitle

\section{Introduction}
\label{sec:introduction}
Due to scientific advancements, currently Machine Learning (ML) is 
widely used in a variety of applications such as image classification, stock predictions, machine translation, and cancer cell detection to name but a few. The benefits of ML comes at a cost of having significant computational overhead, which leads data scientists to turn to Machine Learning-as-a-Service (MLaaS) in order to outsource 
computations. However, the use of MLaaS has raised significant security and privacy concerns.
More precisely, 
a plethora of works 
shows how to successfully attack ML algorithms and gain insight into private data. Examples of such attacks are reconstruction attacks~\cite{zhu2019deep}, model inversion attacks~\cite{hitaj2017deep}, and membership inference attacks~\cite{shokri2017membership}. 
As far as privacy is concerned, ML algorithms are used in fields such as healthcare, personalization and virtual assistants, 
as they can easily leak 
sensitive information related to the users. 
In this survey, we focus on covering existing works, which aim to preserve data privacy in ML algorithms.



\begin{figure}[!ht]
\resizebox{0.5\textwidth}{!}{
	\centering
	\begin{forest}
		for tree={
			line width=1pt,
			if={level()<2}{
				my rounded corners,
				draw=linecol,
			}{},
			edge={color=linecol, >={Triangle[]}, ->},
			if level=0{%
				l sep+=1.3cm,
				align=center,
				parent anchor=south,
				tikz={
					\path (!1.child anchor) coordinate (A) -- () coordinate (B) -- (!l.child anchor) coordinate (C) pic [draw, angle radius=20mm, every node/.append style={fill=white}, "based on"] {angle};
				},
			}{%
				if level=1{%
					parent anchor=south west,
					child anchor=north,
					tier=parting ways,
					align=center,
					font=\bfseries,
					for descendants={
						child anchor=west,
						parent anchor=west,
						anchor=west,
						align=left,
					},
				}{
					if level=2{
						shape=coordinate,
						no edge,
						grow'=0,
						calign with current edge,
						xshift=20pt,
						for descendants={
							parent anchor=south west,
							l sep+=-20pt
						},
						for children={
							edge path={
								\noexpand\path[\forestoption{edge}] (!to tier=parting ways.parent anchor) |- (.child anchor)\forestoption{edge label};
							},
							font=\bfseries,
							for descendants={
								no edge,
							},
						},
					}{},
				},
			}%
		},
		[Privacy-Preserving Techniques
		[Cryptography \\ Methods
		[
		[Homomorphic\\ Encryption ,draw,red
		]
		[Functional \\ Encryption
		]
		[Multiparty \\ Computation ,draw,red
		[]
		]
		[Secure \\ Enclave
		]
		]
		]
		[Data \\ Modification
		[
		[Differential \\ Privacy
		]
		[K-Anonymity
		]
		[Condensation
		]
		[Perturbation
		]
		]
		]
		[Distributed\\ Approaches
		[
		[Federated \\Learning
		[]
		]
		[Split Learning
		[]
		]
		]
		]
		[Hybrid\\ Approaches
		[
		[Cryptographic methods \\ \& Data Modification
		[]
		]
		[Cryptographic methods \\ \& Distributed Approaches
		[]
		]
		[Homomorphic Encryption \\ \& Multi-party Computation,draw,red
		[]
		]    
		]
		]
		]
	\end{forest}
	}
	\caption{Privacy-preserving Techniques. This SoK covers the privacy-preserving techniques marked in red.}
	\label{fig:PPML}
\end{figure}
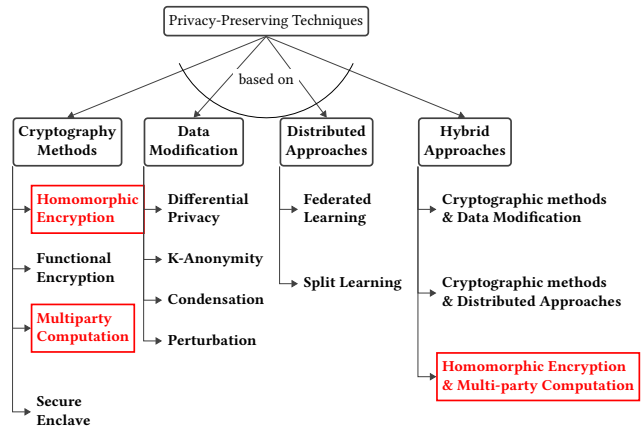

Privacy-Preserving Machine Learning (PPML) is a research field focused on enhancing data privacy in ML using Privacy-Preserving Techniques (PPTs). 
These techniques consist of cryptographic, distributed,
and data modification approaches (\autoref{fig:PPML}). PPML is a highly active line of research, with a significant amount of work in literature~\cite{cabrero2021sok, abuadbba2020can, chamikara2021privacy, tan2021cryptgpu, hesamifard2018privacy, mohassel2017secureml}. The 
main techniques used to achieve PPML are: Homomorphic Encryption (HE)~\cite{gentry2009fully}, Secure Multi-party Computation (SMPC)~\cite{yao1986generate}
, Federated Learning (FL)~\cite{truex2019hybrid}, Differential Privacy (DP)~\cite{dwork2006calibrating}, and Functional Encryption (FE)~\cite{bakas2022private}.
The main aim of this Systemization of Knowledge (SoK) paper is to survey and compare State-of-the-Art (SotA) 
works in the area of HE and SMPC-based PPML in the training phase. 

HE schemes, allow involved parties (users or service providers) to perform computations such as addition, multiplication or bitwise operations on encrypted data. After the computation, the format of the input data remains intact, but its value changes~\cite{acar2018survey}. This property of ``privacy homomorphism'' was initially theorized by Rivest, Adleman, and Dertouzos~\cite{rivest1978data}. The researchers believed that, if the input 
data was homomorphic, 
they would be able to perform computations on encrypted data, similar to how multiplication is performed by the 
RSA public-key cryptosystem. 
As a result, first generation HE schemes, like Paillier~\cite{paillier1999public} were only able to homomorphically compute a single type of mathematical operation. This changed in 2009, when C. Gentry~\cite{gentry2009fully} proposed the first fully HE scheme. Said scheme would allow multiple operations 
on encrypted data. However, because of the computational complexity, various researchers, including Gentry himself, began developing leveled HE schemes~\cite{brakerski2014leveled} or somewhat HE schemes~\cite{fan2012somewhat,pisa2012somewhat} to limit the amount of computations 
per system through different means.

These developments of HE have allowed users to store fully encrypted data in a cloud, while retaining their ability to perform computations on the encrypted data. This minimizes the risks of a malicious service provider or employee eavesdropping on confidential information, that is being stored in their service~\cite{khan2021blind,bakas2022symmetrical, khan2023learning}.
A general example of HE would be, if an encryption scheme would take two separate encrypted inputs $\mathsf{E(a)}$ and $\mathsf{E(b)}$, where $\mathsf{E(\cdot)}$ is an encryption function, and by using an addition or multiplication algorithm it would be able to compute $\mathsf{E(a)} + \mathsf{E(b)} = \mathsf{E(a + b)}$ or $\mathsf{E(a)} \times \mathsf{E(b)} =\mathsf{E(a \times b)}$ respectively, without having to decrypt the inputs~\cite{gentry2009fully}. Aside from being used in cloud services~\cite{POTEY2016175} to protect data, HE can be applied to scenarios such as keeping medical and genome~\cite{wood2020homomorphic,frimpong2024guardml} information private in ML applications.

SMPC aims to create methods that enable different parties to jointly compute a function on their private data, while preserving 
privacy~\cite{Yao1982ProtocolsFS}. 
SMPC was introduced as a 
two-party computation (2PC) by Andrew Yao~\cite{yao1986generate} and was generalized into SMPC by Goldreich, Micali, and Wigderson~\cite{goldreich1987completeness}. 
In 2PC settings~\cite{Yao1982ProtocolsFS}, the two parties jointly compute a function on their inputs without disclosing the values of their inputs to the other party.
The 
2PC setting can be extended to three (3PC)~\cite{araki2017optimized,mohassel2015fast,araki2016high}, 
four (4PC)~\cite{gordon2018secure, byali2018fast,ishai2015secure}, 
or, 
n-party~\cite{zheng2021cerebro,makri2021rabbit,lu2022polymath}
settings, where these parties collaborate to privately compute a joint function of their inputs. 

Said feature has great potential, as it can 
be applied in any situation where sensitive data from two or more parties are computed. Additionally, it could help with many ML applications, 
currently infeasible 
due to privacy concerns. 
An 
application example 
is 
the simplified version of training ML models on private datasets held
by different parties and evaluating one party's private model using another party's private data. Other applications
for SMPC are private DNA comparison, privacy-preserving auctions, and more. To date, 
SMPC has made substantial success in several real-world situations, with a significant payoff to society. 
For instance, it has been used to securely analyze the wage data of~112,600 employees in the greater Boston area in order to quantify pay disparities by gender and race~\cite{varia2018cryptographically}. SMPC has also been used to allow satellite operators to calculate the likelihood of their satellites colliding without having to share their underlying private orbital information~\cite{hemenway2016high}.

\subsection{Motivation \& Contribution}
\label{subsec:contribution}
\noindent \textbf{\textit{Motivation:}} 
ML allows interested parties (e.g.\ companies, researchers, etc.) to train models 
either locally or in an interactive distributed environment. Privacy is guaranteed in the scenario where a model is trained on a local machine, 
as in this case, no models or data are shared. 
However, there are concerns regarding 
the level of security when deploying ML in a distributed environment. For example, in a scenario where parties share their data to train a model. 
To address privacy concerns, various PPML techniques have been proposed. These PPML techniques take both ethical and legal 
concerns into account. Specifically, in privacy-preserving training, data privacy is demanded\footnote{Neither the cloud server nor any other involved parties, learn anything about 
outsourced data of clients other than their size.}. In privacy-preserving MLaaS, model privacy is also necessary besides 
data privacy 
so that neither the client nor any other parties learn anything about the model parameters on cloud servers, other than what can be learned from the final prediction. 

The main motivation behind our study is to better understand the landscape of MLaaS in a data-sensitive context by employing PPML techniques. 
In this study, we focus primarily on the two cryptographic techniques: HE and SMPC -- that are employed in PPML training. The other methods are also employed for PPML 
in various ways, but, 
here, they are not taken into consideration. This decision is backed by a number of reasons 
some of which are covered here. 
SE is a hardware security approach (that uses hardware enclaves) and requires specialized hardware for computation, while HE and SMPC are pure cryptographic approaches and do not require such hardware for computation. As our survey exclusively examines application and 
non-hardware-based approaches, 
SE is not included in this survey. The 
second cryptographic method, FE, has a very constrained range of applications. It has numerous effective constructs for evaluating linear and quadratic functions, but only a few known 
FE schemes 
can be used to evaluate generic functions effectively. Due to the limited number of applications and existing studies, 
FE does not fall within the scope of this work. Furthermore, HE and SMPC are explicitly optimized for a specific notion of privacy while 
other approaches, such as distributed approaches, do help enhance privacy, 
but may not inherently protect privacy and are known to have privacy vulnerabilities~\cite{abuadbba2020can,pasquini2021unleashing}. Hence, they do not form part of 
the scope of this work. Accordingly, while data modification, such as DP, is a common approach of PPML, it too is 
beyond the purview of our study, since 
the techniques we address 
utilize cryptography to hide information, while enabling computations over it. Such techniques also suffer from reducing model utility and are best used in hybrid settings. 

While it is understood that not all research could offer open-source implementations (OSI) that is easy to install and run due to various constraints, doing so would tremendously enhance the value of a published paper. 
As analyzed in~\cite{olszewski2023get} several works in the secure ML domain do not provide any form of OSI or the provided code does not reproduce the expected results\footnote{\cite{olszewski2023get} 
	analyses 
a variety of security related topics in ML, but does not go in to detail of the techniques covered or the results of the covered articles. Our paper analyses the reproducibility of the articles using HE and SMPC in PPML training and analyses their novelty, advantages and limitations.}. So, another key motivation for our work is to examine 
whether existing works in PPML provide support for open science by allowing others to successfully reproduce their results and use the designed models. 
For this, we focus on evaluating systems with OSI and compare 
our results to those reported in the original papers. Furthermore, we base our evaluation on another aspect: the threat model. As our study is centered around two PPTs, namely HE and SMPC, which primarily operate under semi-honest threat model, this choice provides a meaningful framework for assessing these works.


\noindent \textbf{\textit{Contribution:}} In this context 
cover preliminaries for both HE and SMPC techniques, 
as well as examine and compare the libraries used to implement various schemes. We chose~9 SotA papers on HE 
and~17 on SMPC implementation in PPML, compared the methodologies 
and evaluated the outcomes. More specifically: 
\begin{itemize}[leftmargin=0.2cm]
	\item 
 We introduce 
 the intersection of both ML and privacy fields 
 placing special emphasis on cryptographic techniques used to protect the data.
	\item We 
 lay the basic but substantial theoretical foundation that helps researchers comprehend current HE and SMPC-based methods to PPML. We also go through basic ML fundamentals and describe 
 often-used datasets.
	\item We 
 thoroughly review of 
 HE and SMPC-based PPML literature, emphasizing on the strengths and shortcomings of various approaches and assessing how they supplement one another.
	\item We analyze various aspects of the implementations such as 
 security, computational complexity, and adversarial models to get an overview of the current SotA implementations.  We examine the limitations that prevent the existing HE and SMPC-based PPML solutions from being implemented into real-world settings, mostly due to issues with efficiency and usability.
    \item We also highlight the importance of OSI. By encouraging researchers to prioritize OSI, we aim to bridge the gap between theoretical advancements and real-world applications. This promotes improved reproducibility, wider adoption and impact, and long-term sustainability in the field of PPML.
	\item  We lay out future research directions 
 aimed at improving existing works in terms of performance and security. 
\end{itemize}

\noindent \textbf{\textit{Comparison to related surveys:}} Although some similarities are inevitable, our work differs from 
similar works
~\cite{papernot2018sok,lindell2005secure,cabrero2021sok,ng2023sok} in many aspects. In this article, we 
overview 
the entire spectrum of PPML and narrow down our analysis to HE and SMPC 
describing their functionality and limitations in depth. We mainly focus on the training part of DL, where user 
privacy is of critical concern. To further differentiate our work, 
we cover the functionality and performance of SotA PPML approaches and attempt to reproduce the results of OSIs. Aside from that, we compare the 
frameworks covered 
in terms of security, availability and performance 
and provide insight into available libraries. 
We believe that these two last tasks constitute the main contribution of this work. 
This is because these tasks aim to support open science and reproducible research and can give valuable insights to researchers 
using these libraries to conduct further research or to build modern privacy-preserving online services. Since our work is closely related to the work of~\cite{cabrero2021sok} and~\cite{ng2023sok}, we provide a detailed comparison between our work,~\cite{cabrero2021sok} and~\cite{ng2023sok} in~\autoref{tab:SoKcomparisons}.

\begin{table*}
\small
\centering
\caption{Comparison between our paper, ~\cite{cabrero2021sok} and \cite{ng2023sok}}
\label{tab:SoKcomparisons}
\begin{tblr}{
  width = \linewidth,
  colspec = {Q[80]Q[265]Q[270]Q[300]},
   row{1} = {c},
  hlines,
  vlines,
}
 & \cite{cabrero2021sok} & \cite{ng2023sok} & Our paper\\
Focus & Study the landscape of PPML (inference and training) in data-sensitive contexts through HE, SMPC and Hybrid Techniques &  Study 53 papers on privacy-preserving neural network based on HE and SMPC &   Study the landscape of PPML training as a service using HE and SMPC and Hybrid Techniques\\
Methodology & {
\labelitemi   Problem addressed, training or inference\\\labelitemi   The architecture proposed, i.e., centralized, distributed, or hybrid\\\labelitemi   Privacy goals and adversarial model\\\labelitemi   The particular techniques involved, i.e., SMPC, HE and/or others\\\labelitemi   The issues considered regarding efficiency and usability.} & There is not a clear framework on how the paper studies each work, but rather it~systematizes notable works based on their use of cryptographic primitives and tricks, hence point out the relationships between these works & {
\labelitemi Threat model, supported layers and evaluation datasets \\\labelitemi Cryptographic technique used (e.g. HE encryption schemes, or SMPC cryptographic primitives)\\\labelitemi          Explores networking type and NN architecture in SMPC approaches to better evaluate the testing environments and results \\\labelitemi      Reviews HE settings and protocol design trends.     \\\labelitemi      Perform experiments using OSI of the selected approaches.}\\
PPTs & HE, SMPC, DP when it intersects with HE / SMPC & Only HE and SMPC &  Only HE and SMPC\\
Categorization of surveyed works &  No categorization for works in privacy-preserving training & Categorizing works based on cryptographic primitives and techniques for both HE and SMPC & {\labelitemi  HE: Categorizing works based on HE techniques\\\labelitemi SPMC: Categorizing works based on the number of parties involved in the privacy-preserving training protocol two, three, four-party computation)}\\
PPML works & 2 for HE and 5 for SMPC & Study 12 works that support private training &  9 for HE and 17 for SMPC\\
Reproducing results & No & Focus on private inference and 4 private training results reported & Focus on private training phase\\
Pros & Cover both training and inference phase & {\labelitemi  Cover both training and inference phase\\\labelitemi Evaluation of many works on the MNIST and CIFAR dataset in the private inference phase\\\labelitemi Identifying various challenges and open problems in current cryptographic NN computation} & {\labelitemi  Cover in-depth the privacy-preserving training phase using HE and  SMPC \\\labelitemi Contain the evaluation of the datasets and OSI of the surveyed works in the private training phase\\\labelitemi Identifying various challenges and reproducibility problems regarding secure NN training}\\
Cons & {\labelitemi  Quite brief in the survey of works in PPML training phase\\\labelitemi   No categorizations for PPML training works\\\labelitemi   No evaluation of datasets or implementations} &  Only briefly study the private-training phase &  
Does not cover the inference phase
\end{tblr}
\end{table*}

\subsection{Organization}
\label{subsec: organization}
The rest of paper is organized as follows. 
In~\autoref{sec:scope}, we define the scope and methodology. 
We overview the primary SotA HE and SMPC implementations of the SoK in~\autoref{sec:SotA}, more specifically we talk about how they are implemented, what datasets 
and which schemes are used. This section also compares the documented results of the research papers with other implementations mentioned in the original paper and provides an overview of the results to compare each implementation to one another. Then ~\autoref{section:evaluation} covers the experiments we conducted for both HE and SMPC implementations and the results we received from our tests. This section also covers various aspects of the HE and SMPC protocols such as the libraries used and how they differ, as well as the security, computational complexity and adversarial model of the proposed HE and SMPC implementations. The main takeaways from the survey are 
provided in~\autoref{sec:takeaways}, followed by challenges and future directions in~\autoref{sec:challenges}, and 
conclude the paper in~\autoref{sec:conclusion}. In~\autoref{sec: preliminaries}, we overview the categorization, history and schemes of HE and SMPC and also the formal definitions for both. We also 
refer to some of the libraries used to realize these schemes into various implementations for secure and private storage and computations. 

\section{Scope and Methodology}    
\label{sec:scope}

Cryptographic techniques encounter 
difficulties when used with ML. For example, although HE can relieve the client endpoint of a significant workload, it can only compute a limited number of operations, when dealing with complex problems (due to performance issues). In addition, using HE for ML 
can be complicated, as 
some ML operations, such as non-linear activation functions, are incompatible with most HE, except TFHE~\cite{chillotti2020tfhe}, and can be simulated with 
polynomial approximation. 
As such, the main concerns 
when dealing with PPML using HE, 
relate to efficiency and usability. PPML can also be deployed in a collaborative setting using SMPC, where different stakeholders contribute with their data to a common goal. The main concern using SMPC is the possibility of malicious behavior by an external entity or even by a subset of participating parties as well as the high communication overhead. 

It is important to research and assess various aspects of both HE and SMPC 
to better understand their effectiveness and usability, as 
both of the selected approaches have different advantages and disadvantages. 
The paper studies various aspects 
of each of the proposed approaches, such as the threat model, the supported layers, the corresponding techniques, 
and the evaluation datasets. As for SMPC, we cover the networking type 
and the NN architecture adopted in 
said approaches to better evaluate the testing environments and results. For HE, we review the HE setting to gain knowledge on current protocol design trends and the differences 
that occur depending on the setting chosen. To approach 
these aspects we overview 
relevant literature and perform experiments on 
available OSIs of the selected approaches. Each aspect contains information 
that helps analyze the proposed approaches in terms of both 
applicability and security. Analyzing the threat model allows us to gain a better understanding of the attacker's abilities 
when 
faced with the protocol. 
This in turn describes the potential security of the protocol. Aspects such as the supported layers and evaluation datasets show the applicability of the protocol in real-world applications, as modern solutions require relatively large datasets for proper training and various layers for different task applications. Aside from theoretical aspects, we also analyze various experimental results, which are important for the selected method, such as the training time and accuracy for both methods, the communication overhead for SMPC approaches, and the testing environment specifications and complexity for HE approaches. 

To analyse 
the aforementioned aspects in SMPC and HE PPML implementations, we aimed to 
identify 
literature 
encapsulating them in 
experiments or methodology. The PPML implementations were identified by analysing works published in ML or information security-focused venues and by cross-checking the referenced works in those papers. Our main focus was to identify and categorise literature 
using different libraries 
trained and tested on 
various datasets, and 
implementing varied techniques, schemes and settings. As such, we identified 9 HE-based PPML approaches 
covering the 
various schemes and settings in the HE domain and 17 SMPC-based PPML implementations. To the best of our knowledge, these 26 works show the primary areas of focus in cryptographic PPML, by encapsulating the different aspects covered previously. 

\section{State-of-the-Art Approaches}
\label{sec:SotA}

\subsection{Secure Training using HE}
\label{subsec:HEProtoc}

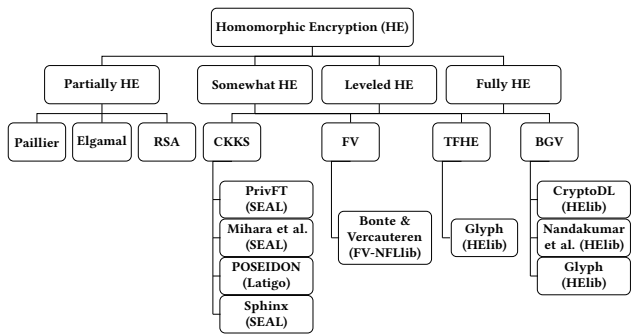
\begin{figure}[!h]
\resizebox{0.47\textwidth}{!}{
	\centering
    \large
	\begin{tikzpicture}[node distance=1.5cm]
		\node (HE) [roundrect] { \textbf{Homomorphic Encryption (HE)}};
		\node (PHE) [roundrect, below of = HE,xshift=-5.5cm] {\textbf{Partially HE}};
		\node (SHE) [roundrect, below of = HE,xshift=-1.5cm] {\textbf{Somewhat HE}};
		\node (LHE) [roundrect, below of = HE,xshift=1.75cm] {\textbf{Leveled HE}};
		\node (FHE) [roundrect, below of = HE,xshift=5cm] {\textbf{Fully HE}};
		\node (CKKS) [roundrect2, below of = SHE,xshift=-0.6cm] {\textbf{CKKS}};
		\node (FV) [roundrect2, right of = CKKS,xshift=1.6cm] {\textbf{FV}};
		\node (TFHE) [roundrect2, right of = FV,xshift=1.4cm] {\textbf{TFHE}};
		\node (BGV) [roundrect2, right of = TFHE,xshift=0.8cm] {\textbf{BGV}};
		\node (Pail) [roundrect2, below of = PHE,xshift=-1.7cm] {\textbf{Paillier}};
		\node (Elg) [roundrect2, right of = Pail,xshift=0.2cm] {\textbf{Elgamal}};
		\node (RSA) [roundrect2, right of = Elg,xshift=0.2cm] {\textbf{RSA}};
		\draw (HE.south) -- ++(0,-0.25cm) coordinate (div);
		\draw (div) -| (PHE.north);
		\draw (div) -| (SHE.north);
		\draw (div) -| (LHE.north);
		\draw (div) -| (FHE.north);
		\draw (PHE.south) -- ++(0,-0.25cm) coordinate (divP);
		\draw (divP) -| (Pail.north);
		\draw (divP) -| (Elg.north);
		\draw (divP) -| (RSA.north);
		\draw (SHE.south) -- ++(0,-0.25cm) coordinate (divS);
		\draw (LHE.south) -- ++(0,-0.25cm) coordinate (divL);
		\draw (FHE.south) -- ++(0,-0.25cm) coordinate (divF);
		\draw (divS) -| (CKKS.north);
		\draw (divS) -| (FV.north);
		\draw (divS) -| (TFHE.north);
		\draw (divS) -| (BGV.north);
		\draw (CKKS.south) -| ++(-0.5cm,-1cm) coordinate (divPriv);
		\draw (divPriv) -| ++(0.2cm,0cm) coordinate (divPriv2);
		\draw (divPriv) -| ++(0cm,-1cm) coordinate (divMiha);
		\node (Priv) [roundrect3, right of = divPriv, xshift = -0.1cm, text width = 2.2cm, text centered] {\textbf{PrivFT\\ (SEAL)}};
		\draw (divMiha) -| ++(0.2cm,0cm) coordinate (divMiha2);
		\draw (divMiha) -| ++(0cm,-1cm) coordinate (divPose);
		\node (Miha) [roundrect3, right of = divMiha, xshift = -0.1cm, text width = 2.2cm, text centered] {\textbf{Mihara et al. \\ (SEAL)}};
		\draw (divPose) -| ++(0.2cm,0cm) coordinate (divPose2);
		\node (Pose) [roundrect3, right of = divPose, xshift = -0.1cm, text width = 2.2cm, text centered] {\textbf{POSEIDON (Latigo)}};
        \draw (divPose) -| ++(0cm,-1cm) coordinate (divSphinx);
        \draw (divSphinx) -| ++(0.2cm,0cm) coordinate (divSphinx2);
		\node (Sphinx) [roundrect3, right of = divSphinx, xshift = -0.1cm, text width = 2.2cm, text centered] {\textbf{Sphinx\\ (SEAL)}};
		\draw (FV.south) -| ++(-0.5cm,-2cm) coordinate (divBV);
		\draw (divBV) -| ++(0.2cm,0cm) coordinate (divBV2);
		\node (BV) [roundrect3, right of = divBV, xshift = -0.1cm, text width = 2.2cm, text centered] {\textbf{Bonte \& Vercauteren (FV-NFLlib)}};
		\draw (TFHE.south) -| ++(-0.5cm,-2cm) coordinate (divGlyph);
		\draw (divGlyph) -| ++(0.2cm,0cm) coordinate (divGlyph2);
		\node (Glyph) [roundrect3, right of = divGlyph, xshift=-0.4cm, text width = 1.5cm, text centered] {\textbf{Glyph (HElib)}};
		\draw (BGV.south) -| ++(-0.5cm,-1cm) coordinate(divCrypto);
		\draw (divCrypto) -| ++(0.2cm,0cm) coordinate (divCrypto2);
		\draw (divCrypto) -| ++(0cm,-1cm) coordinate (divNanda);
		\node (Crypto) [roundrect3, right of = divCrypto, xshift = -0.1cm, text width = 2.2cm, text centered] {\textbf{CryptoDL\\ (HElib)}};
		\draw (divNanda) -| ++(0.2cm,0cm) coordinate (divNanda2);
		\draw (divNanda) -| ++(0cm,-1cm) coordinate (divGlyphB);
		\node (Nanda) [roundrect3, right of = divNanda, xshift = -0.1cm, text width = 2.2cm, text centered] {\textbf{Nandakumar et al. (HElib)}};
		\draw (divGlyphB) -| ++(0.2cm,0cm) coordinate (divGlyphB2);
		\node (Glyph2) [roundrect3, right of = divGlyphB, xshift = -0.1cm, text width = 2.2cm, text centered] {\textbf{Glyph (HElib)}};
	\end{tikzpicture}
	}
	\caption{HE Taxonomy}
	\label{fig:hetaxonomy}
\end{figure}

As covered in the preliminaries, 
implementation of HE in PPML models depends on 
mathematical operations that the HE scheme can use and on whether the HE technique works on binary~\cite{chillotti2016faster}, integer~\cite{fan2012somewhat} or approximate numbers~\cite{cheon2017homomorphic} (see~\autoref{table: hecomparison}\footnote{Due to space constraints, all the tables are moved to the appendix section.}). The overall performance and cost of the Neural Network (NN) will vary depending on the HE scheme selected. There is no single HE approach that outperforms all 
others; 
rather, performance varies from application to application depending on the HE scheme selected~\cite{viand2021sok}. 
To 
further clarify the implementations covered, 
we have categorized them according to the underlined HE scheme (see~\autoref{fig:hetaxonomy}). 


\noindent\textbf{FV-based implementations:} Bonte and Vercauteren~\cite{bonte2018privacy} -- is a Logistic Regression (LoR)~\cite{kleinbaum2002logistic} model based on the SHE implementation of the FV scheme. The implementation uses the fixed Hessian method at a low depth to construct an HE algorithm capable of privacy-preserving logistic training. To accomplish this, the researchers show that a practical algorithm can be constructed by using the Simplified Fixed Hessian (SFH) method. 
When compared 
to Matlab's \texttt{glmfit} function~\cite{mathworks2022glmfit}, 
the SFH algorithm produces equivalent accuracy, 
while providing superior security for the data. 
Alongside FV schemes there is 
BGV, another HE scheme 
allowing performing homomorphic operations on integers arithmetic. The core difference between them 
is their plaintext encoding. 
Namely, BGV encodes the messages starting from the least significant bit, while FV 
starts from the most significant bit.

\noindent\textbf{BGV-based implementations:} CryptoDL~\cite{hesamifard2018privacy} -- is an SHE implementation 
that uses the BGV scheme to encrypt data. The open-source library HElib~\cite{HELib} 
is used for implementing the BGV scheme and 
also contains various optimizations for HE.  CryptoDL is one of the first implementations, 
providing a solution for NN training using encrypted training data through HE. In their work, the researchers show how to 
locate approximations of low-degree polynomials and how to approximate non-linear activation functions, such as ReLU, sigmoid, and tanh. The proposed implementation was trained and tested on various datasets, such as CIFAR~\cite{krizhevsky2009learning}, MNIST~\cite{lecun1998mnist}, and UCI~\cite{almeida_silva_santos_hidalgo_2017}. The research is compared to CryptoNets~\cite{gilad2016cryptonets} (an HE implementation of NN classification) and SecureML~\cite{chen2020parsecureml}. CryptoDL~\cite{hesamifard2018privacy} outperforms the other implementations both in terms of accuracy, 
classification throughput, 
and communication overhead. However, it only states a training time of~10476.29 seconds for MNIST implementation and doesn't provide further comparisons with other implementations.


Nandakumar \textit{et al.}~\cite{nandakumar2019towards} -- is an FHE implementation of PPML training. 
The model is trained in a non-interactive way and utilizes the BGV scheme through HElib. 
The research improves on prior implementations by 
optimizing the way ciphertext is packed to reduce bootstrapping and 
encourage 
parallel training. The implementation uses the MNIST dataset for training and classification and protects the client privacy 
by encrypting both the training data and the model parameters. 
According to the research results, the implementation reaches~96.4\% and~97.8\% accuracy on different NN architectures, when training for~50 epochs on the plaintext, while the computation time for a single mini-batch of~60 training samples varies between~40 minutes and~9.4 hours with 
their optimized packing and parallelization. 

Research on the BGV scheme led to Cheon \textit{et al.} to propose a major improvement to the scheme 
expanding the homomorphic operation to floating point arithmetic in the CKKS scheme~\cite{cheon2017homomorphic}. This scheme has greatly impacted the development of new PPML works 
and is used by most SoTA works.

%

\noindent\textbf{CKKS-based implementations:} PrivFT~\cite{al2020privft} -- is an LHE implementation based on the CKKS LHE scheme. 
It allows the use of a GPU for faster training 
and faster inference. 
The GPU implementation is achieved by using the residual number system (variant of the CKKS scheme) and by using CUDA~10~\cite{Nvidia:CUDA}. 
The implementation was tested on various datasets, such as Yelp Dataset~\cite{inc._2018}, AGNews news topic classification~\cite{gulli_ferragina_2005}, IMDB movie reviews~\cite{maas-EtAl:2011:ACL-HLT2011,LargeMovie:Dataset} and DBPedia ontology classification~\cite{community_2021}, to name but a few. PrivFT compared the accuracy of the model with XLNet~\cite{yang2019xlnet}, BERT ITPT~\cite{sun2019fine} and ULMFit~\cite{howard2018universal} on~4~datasets (Yelp, AGNews, IMDB and DBPedia) and noted that the accuracy of PrivFT was lower than the other models. The largest accuracy difference noted was on the IMDB dataset, when compared to the XLNet model~(91.49\% 
versus~96.21\%). However, the main advantage of PrivFT is its faster training, boosted by GPU usage, which accelerates training by approximately~2.2x.

Sphinx~\cite{tian2022sphinx} -- is a CKKS LHE scheme used for online training and inference on the cloud. The code is implemented with the Microsoft SEAL library for homomorphic operations and makes use of KANN\footnote{\href{https://github.com/attractivechaos/kann}{https://github.com/attractivechaos/kann}} to implement ML models in C. Sphinx combines various improvements, such as batch packing, made in HE PPML training and inference to greatly reduce the communication overhead and computational complexity of ML operations. The authors improve HE multiplications by reducing the amount of rescaling and relinearization operations. 
They also 
use 
forward propagation caching 
and introduce a different encryption method, called zero encryption, to reduce communication costs. The implementation is tested on MNIST and CIFAR-10 datasets and shows reduced communication costs and computational complexity when compared to baseline HE. 
It reports higher throughput and lower latency than works such as SecureNN~\cite{wagh2019securenn}, SecureML~\cite{mohassel2017secureml} and CryptoDL~\cite{hesamifard2018privacy}.


Mihara \textit{et al.}~\cite{mihara2020neural} -- is a CKKS-based LHE implementation 
using the Microsoft SEAL library. 
The main optimization 
of this implementation is that it provides a novel weight matrix packing method,  
increasing the speed of the training phase without losing accuracy during the inference phase. The packing method in question packs weights in a matrix diagonally instead of in a row, which 
keeps the amount of operations 
low as the packing does not require multiplication. This 
in turn reduces the complexity of the circuit. 
The authors 
note that their packing method reduces the time of training for one iteration from~28.47s (row packing) to~9.25s (diagonal packing). 
They test 
and compare 
their implementation with the same architecture plaintext NN 
and receive 
similar results~(98.05\% accuracy) to their ciphertext model~(98.47\%).

POSEIDON~\cite{sav2020poseidon} -- is a hybrid PPML implementation that makes use of federated learning and LHE techniques to produce Multiparty Homomorphic Encryption (MHE). POSEIDON's MHE implementation 
uses the CKKS scheme and is an extended version of the Mouchet \textit{et al.}~\cite{MouchetTroncoso-PastorizaBossuatHubaux+2021+291+311} implementation. The implementation provides confidentiality for: \begin{inparaenum}[\it (i)] \item Training data, \item Model details and \item Evaluation data\end{inparaenum}. POSEIDON's 
accuracy and training time is similar to other SotA SMPC techniques, such as SecureML~\cite{mohassel2017secureml}, SecureNN~\cite{wagh2019securenn} and Falcon~\cite{wagh2021falcon}. 
When compared to other HE techniques, like Nandakumar \textit{et al.}~\cite{nandakumar2019towards} and CryptoDL~\cite{hesamifard2018privacy}, it outperforms them on a relatively large margin.

The core limitation of FV, BGV and CKKS works 
is that these schemes have very expensive and impractical bootstrapping operations 
resulting to increased computational complexity. To address this Chillotti \textit{et al.}~\cite{chillotti2016faster} proposed TFHE, a scheme 
that speeds up bootstrapping operations.



%

\noindent\textbf{TFHE-based implementations:} Glyph~\cite{lou2020glyph} -- is a hybrid FHE approach, 
using two different cryptosystems, namely TFHE and BGV. TFHE is used to implement non-linear activation functions and BGV scheme is used to perform multiply-accumulate operations. Another benefit, 
is that switching between the two cryptosystems 
results in significantly reduced training times compared to other PPML techniques. 
In the results, the researchers showed that the accuracy of their implementation was comparable 
to Chimera~\cite{boura2020chimera} -- another hybrid HE technique. Chimera 
is similar to Glyph as it switches between the TFHE and BFV. 
An additional benefit of Glyph 
is that the training time~(8 days) for the MNIST dataset is~2.6$\times$ faster than the Chimera implementation (28.6 days). 
Furthermore, Glyph was 
slightly more accurate than Chimera (98.6\% vs~97.8\%). The researchers also compared Glyph with the Nandakumar \textit{et al.}~\cite{nandakumar2019towards} implementation and showed that Glyph was significantly faster during training (8 days compared to~13.4 years).

\subsection{Secure Training using SMPC}
\label{subsec: secMPC}

\begin{figure}[!h]
\resizebox{0.48\textwidth}{!}{
	\centering
\tikzset{every picture/.style={line width=0.75pt}} 

\begin{tikzpicture}[x=0.75pt,y=0.75pt,yscale=-1,xscale=1]

\draw  [color={rgb, 255:red, 74; green, 74; blue, 74 }  ,draw opacity=1 ] (118,124.8) .. controls (118,122.15) and (120.15,120) .. (122.8,120) -- (165.2,120) .. controls (167.85,120) and (170,122.15) .. (170,124.8) -- (170,139.2) .. controls (170,141.85) and (167.85,144) .. (165.2,144) -- (122.8,144) .. controls (120.15,144) and (118,141.85) .. (118,139.2) -- cycle ;
\draw  [fill={rgb, 255:red, 155; green, 155; blue, 155 }  ,fill opacity=1 ][dash pattern={on 0.84pt off 2.51pt}] (116.3,144) .. controls (116.3,139.58) and (119.21,136) .. (122.8,136) .. controls (126.39,136) and (129.3,139.58) .. (129.3,144) .. controls (129.3,148.42) and (126.39,152) .. (122.8,152) .. controls (119.21,152) and (116.3,148.42) .. (116.3,144) -- cycle ;
\draw  [color={rgb, 255:red, 74; green, 74; blue, 74 }  ,draw opacity=1 ] (201,65.8) .. controls (201,63.15) and (203.15,61) .. (205.8,61) -- (245.2,61) .. controls (247.85,61) and (250,63.15) .. (250,65.8) -- (250,80.2) .. controls (250,82.85) and (247.85,85) .. (245.2,85) -- (205.8,85) .. controls (203.15,85) and (201,82.85) .. (201,80.2) -- cycle ;
\draw  [color={rgb, 255:red, 74; green, 74; blue, 74 }  ,draw opacity=1 ][fill={rgb, 255:red, 155; green, 155; blue, 155 }  ,fill opacity=1 ][dash pattern={on 0.84pt off 2.51pt}] (197,84) .. controls (197,79.58) and (199.91,76) .. (203.5,76) .. controls (207.09,76) and (210,79.58) .. (210,84) .. controls (210,88.42) and (207.09,92) .. (203.5,92) .. controls (199.91,92) and (197,88.42) .. (197,84) -- cycle ;
\draw  [color={rgb, 255:red, 74; green, 74; blue, 74 }  ,draw opacity=1 ] (200.8,173.07) .. controls (200.8,170.42) and (202.95,168.27) .. (205.6,168.27) -- (246.2,168.27) .. controls (248.85,168.27) and (251,170.42) .. (251,173.07) -- (251,187.47) .. controls (251,190.12) and (248.85,192.27) .. (246.2,192.27) -- (205.6,192.27) .. controls (202.95,192.27) and (200.8,190.12) .. (200.8,187.47) -- cycle ;
\draw  [color={rgb, 255:red, 74; green, 74; blue, 74 }  ,draw opacity=1 ][fill={rgb, 255:red, 155; green, 155; blue, 155 }  ,fill opacity=1 ][dash pattern={on 0.84pt off 2.51pt}] (196.8,191.6) .. controls (196.8,187.18) and (199.71,183.6) .. (203.3,183.6) .. controls (206.89,183.6) and (209.8,187.18) .. (209.8,191.6) .. controls (209.8,196.02) and (206.89,199.6) .. (203.3,199.6) .. controls (199.71,199.6) and (196.8,196.02) .. (196.8,191.6) -- cycle ;
\draw  [color={rgb, 255:red, 74; green, 74; blue, 74 }  ,draw opacity=1 ] (200.6,213.8) .. controls (200.6,211.15) and (202.75,209) .. (205.4,209) -- (246.2,209) .. controls (248.85,209) and (251,211.15) .. (251,213.8) -- (251,228.2) .. controls (251,230.85) and (248.85,233) .. (246.2,233) -- (205.4,233) .. controls (202.75,233) and (200.6,230.85) .. (200.6,228.2) -- cycle ;
\draw  [color={rgb, 255:red, 74; green, 74; blue, 74 }  ,draw opacity=1 ][fill={rgb, 255:red, 155; green, 155; blue, 155 }  ,fill opacity=1 ][dash pattern={on 0.84pt off 2.51pt}] (196.6,232) .. controls (196.6,227.58) and (199.51,224) .. (203.1,224) .. controls (206.69,224) and (209.6,227.58) .. (209.6,232) .. controls (209.6,236.42) and (206.69,240) .. (203.1,240) .. controls (199.51,240) and (196.6,236.42) .. (196.6,232) -- cycle ;
\draw  [color={rgb, 255:red, 74; green, 74; blue, 74 }  ,draw opacity=1 ] (201,104.4) .. controls (201,101.75) and (203.15,99.6) .. (205.8,99.6) -- (246.53,99.6) .. controls (249.18,99.6) and (251.33,101.75) .. (251.33,104.4) -- (251.33,118.8) .. controls (251.33,121.45) and (249.18,123.6) .. (246.53,123.6) -- (205.8,123.6) .. controls (203.15,123.6) and (201,121.45) .. (201,118.8) -- cycle ;
\draw  [color={rgb, 255:red, 74; green, 74; blue, 74 }  ,draw opacity=1 ][fill={rgb, 255:red, 155; green, 155; blue, 155 }  ,fill opacity=1 ][dash pattern={on 0.84pt off 2.51pt}] (197,122.6) .. controls (197,118.18) and (199.91,114.6) .. (203.5,114.6) .. controls (207.09,114.6) and (210,118.18) .. (210,122.6) .. controls (210,127.02) and (207.09,130.6) .. (203.5,130.6) .. controls (199.91,130.6) and (197,127.02) .. (197,122.6) -- cycle ;
\draw  [color={rgb, 255:red, 74; green, 74; blue, 74 }  ,draw opacity=1 ] (292.2,105) .. controls (292.2,102.35) and (294.35,100.2) .. (297,100.2) -- (336.2,100.2) .. controls (338.85,100.2) and (341,102.35) .. (341,105) -- (341,119.4) .. controls (341,122.05) and (338.85,124.2) .. (336.2,124.2) -- (297,124.2) .. controls (294.35,124.2) and (292.2,122.05) .. (292.2,119.4) -- cycle ;
\draw  [color={rgb, 255:red, 74; green, 74; blue, 74 }  ,draw opacity=1 ][fill={rgb, 255:red, 155; green, 155; blue, 155 }  ,fill opacity=1 ][dash pattern={on 0.84pt off 2.51pt}] (288.2,123.2) .. controls (288.2,118.78) and (291.11,115.2) .. (294.7,115.2) .. controls (298.29,115.2) and (301.2,118.78) .. (301.2,123.2) .. controls (301.2,127.62) and (298.29,131.2) .. (294.7,131.2) .. controls (291.11,131.2) and (288.2,127.62) .. (288.2,123.2) -- cycle ;
\draw  [color={rgb, 255:red, 74; green, 74; blue, 74 }  ,draw opacity=1 ] (291,146.2) .. controls (291,143.55) and (293.15,141.4) .. (295.8,141.4) -- (334.87,141.4) .. controls (337.52,141.4) and (339.67,143.55) .. (339.67,146.2) -- (339.67,160.6) .. controls (339.67,163.25) and (337.52,165.4) .. (334.87,165.4) -- (295.8,165.4) .. controls (293.15,165.4) and (291,163.25) .. (291,160.6) -- cycle ;
\draw  [color={rgb, 255:red, 74; green, 74; blue, 74 }  ,draw opacity=1 ][fill={rgb, 255:red, 155; green, 155; blue, 155 }  ,fill opacity=1 ][dash pattern={on 0.84pt off 2.51pt}] (287,164.4) .. controls (287,159.98) and (289.91,156.4) .. (293.5,156.4) .. controls (297.09,156.4) and (300,159.98) .. (300,164.4) .. controls (300,168.82) and (297.09,172.4) .. (293.5,172.4) .. controls (289.91,172.4) and (287,168.82) .. (287,164.4) -- cycle ;
\draw  [color={rgb, 255:red, 74; green, 74; blue, 74 }  ,draw opacity=1 ] (290.6,185.8) .. controls (290.6,183.15) and (292.75,181) .. (295.4,181) -- (334.87,181) .. controls (337.52,181) and (339.67,183.15) .. (339.67,185.8) -- (339.67,200.2) .. controls (339.67,202.85) and (337.52,205) .. (334.87,205) -- (295.4,205) .. controls (292.75,205) and (290.6,202.85) .. (290.6,200.2) -- cycle ;
\draw  [color={rgb, 255:red, 74; green, 74; blue, 74 }  ,draw opacity=1 ][fill={rgb, 255:red, 155; green, 155; blue, 155 }  ,fill opacity=1 ][dash pattern={on 0.84pt off 2.51pt}] (286.6,204) .. controls (286.6,199.58) and (289.51,196) .. (293.1,196) .. controls (296.69,196) and (299.6,199.58) .. (299.6,204) .. controls (299.6,208.42) and (296.69,212) .. (293.1,212) .. controls (289.51,212) and (286.6,208.42) .. (286.6,204) -- cycle ;
\draw  [color={rgb, 255:red, 74; green, 74; blue, 74 }  ,draw opacity=1 ] (291.4,66.2) .. controls (291.4,63.55) and (293.55,61.4) .. (296.2,61.4) -- (335.2,61.4) .. controls (337.85,61.4) and (340,63.55) .. (340,66.2) -- (340,80.6) .. controls (340,83.25) and (337.85,85.4) .. (335.2,85.4) -- (296.2,85.4) .. controls (293.55,85.4) and (291.4,83.25) .. (291.4,80.6) -- cycle ;
\draw  [color={rgb, 255:red, 74; green, 74; blue, 74 }  ,draw opacity=1 ][fill={rgb, 255:red, 155; green, 155; blue, 155 }  ,fill opacity=1 ][dash pattern={on 0.84pt off 2.51pt}] (287.4,84.4) .. controls (287.4,79.98) and (290.31,76.4) .. (293.9,76.4) .. controls (297.49,76.4) and (300.4,79.98) .. (300.4,84.4) .. controls (300.4,88.82) and (297.49,92.4) .. (293.9,92.4) .. controls (290.31,92.4) and (287.4,88.82) .. (287.4,84.4) -- cycle ;
\draw  [color={rgb, 255:red, 74; green, 74; blue, 74 }  ,draw opacity=1 ] (368,185) .. controls (368,182.35) and (370.15,180.2) .. (372.8,180.2) -- (416.2,180.2) .. controls (418.85,180.2) and (421,182.35) .. (421,185) -- (421,199.4) .. controls (421,202.05) and (418.85,204.2) .. (416.2,204.2) -- (372.8,204.2) .. controls (370.15,204.2) and (368,202.05) .. (368,199.4) -- cycle ;
\draw  [color={rgb, 255:red, 74; green, 74; blue, 74 }  ,draw opacity=1 ][fill={rgb, 255:red, 155; green, 155; blue, 155 }  ,fill opacity=1 ][dash pattern={on 0.84pt off 2.51pt}] (364,203.2) .. controls (364,198.78) and (366.91,195.2) .. (370.5,195.2) .. controls (374.09,195.2) and (377,198.78) .. (377,203.2) .. controls (377,207.62) and (374.09,211.2) .. (370.5,211.2) .. controls (366.91,211.2) and (364,207.62) .. (364,203.2) -- cycle ;
\draw  [color={rgb, 255:red, 74; green, 74; blue, 74 }  ,draw opacity=1 ] (367.2,145.8) .. controls (367.2,143.15) and (369.35,141) .. (372,141) -- (415.87,141) .. controls (418.52,141) and (420.67,143.15) .. (420.67,145.8) -- (420.67,160.2) .. controls (420.67,162.85) and (418.52,165) .. (415.87,165) -- (372,165) .. controls (369.35,165) and (367.2,162.85) .. (367.2,160.2) -- cycle ;
\draw  [color={rgb, 255:red, 74; green, 74; blue, 74 }  ,draw opacity=1 ][fill={rgb, 255:red, 155; green, 155; blue, 155 }  ,fill opacity=1 ][dash pattern={on 0.84pt off 2.51pt}] (363.2,164) .. controls (363.2,159.58) and (366.11,156) .. (369.7,156) .. controls (373.29,156) and (376.2,159.58) .. (376.2,164) .. controls (376.2,168.42) and (373.29,172) .. (369.7,172) .. controls (366.11,172) and (363.2,168.42) .. (363.2,164) -- cycle ;
\draw  [color={rgb, 255:red, 74; green, 74; blue, 74 }  ,draw opacity=1 ] (366.8,104.2) .. controls (366.8,101.55) and (368.95,99.4) .. (371.6,99.4) -- (414.87,99.4) .. controls (417.52,99.4) and (419.67,101.55) .. (419.67,104.2) -- (419.67,118.6) .. controls (419.67,121.25) and (417.52,123.4) .. (414.87,123.4) -- (371.6,123.4) .. controls (368.95,123.4) and (366.8,121.25) .. (366.8,118.6) -- cycle ;
\draw  [color={rgb, 255:red, 74; green, 74; blue, 74 }  ,draw opacity=1 ][fill={rgb, 255:red, 155; green, 155; blue, 155 }  ,fill opacity=1 ][dash pattern={on 0.84pt off 2.51pt}] (362.8,122.4) .. controls (362.8,117.98) and (365.71,114.4) .. (369.3,114.4) .. controls (372.89,114.4) and (375.8,117.98) .. (375.8,122.4) .. controls (375.8,126.82) and (372.89,130.4) .. (369.3,130.4) .. controls (365.71,130.4) and (362.8,126.82) .. (362.8,122.4) -- cycle ;
\draw  [color={rgb, 255:red, 74; green, 74; blue, 74 }  ,draw opacity=1 ] (444.13,104.07) .. controls (444.13,101.42) and (446.28,99.27) .. (448.93,99.27) -- (495.53,99.27) .. controls (498.18,99.27) and (500.33,101.42) .. (500.33,104.07) -- (500.33,118.47) .. controls (500.33,121.12) and (498.18,123.27) .. (495.53,123.27) -- (448.93,123.27) .. controls (446.28,123.27) and (444.13,121.12) .. (444.13,118.47) -- cycle ;
\draw  [fill={rgb, 255:red, 155; green, 155; blue, 155 }  ,fill opacity=1 ][dash pattern={on 0.84pt off 2.51pt}] (440.13,122.27) .. controls (440.13,117.85) and (443.04,114.27) .. (446.63,114.27) .. controls (450.22,114.27) and (453.13,117.85) .. (453.13,122.27) .. controls (453.13,126.68) and (450.22,130.27) .. (446.63,130.27) .. controls (443.04,130.27) and (440.13,126.68) .. (440.13,122.27) -- cycle ;
\draw  [color={rgb, 255:red, 74; green, 74; blue, 74 }  ,draw opacity=1 ] (367.6,65.8) .. controls (367.6,63.15) and (369.75,61) .. (372.4,61) -- (415.2,61) .. controls (417.85,61) and (420,63.15) .. (420,65.8) -- (420,80.2) .. controls (420,82.85) and (417.85,85) .. (415.2,85) -- (372.4,85) .. controls (369.75,85) and (367.6,82.85) .. (367.6,80.2) -- cycle ;
\draw  [color={rgb, 255:red, 74; green, 74; blue, 74 }  ,draw opacity=1 ][fill={rgb, 255:red, 155; green, 155; blue, 155 }  ,fill opacity=1 ][dash pattern={on 0.84pt off 2.51pt}] (363.6,84) .. controls (363.6,79.58) and (366.51,76) .. (370.1,76) .. controls (373.69,76) and (376.6,79.58) .. (376.6,84) .. controls (376.6,88.42) and (373.69,92) .. (370.1,92) .. controls (366.51,92) and (363.6,88.42) .. (363.6,84) -- cycle ;
\draw  [color={rgb, 255:red, 74; green, 74; blue, 74 }  ,draw opacity=1 ] (367.47,27.33) .. controls (367.47,24.68) and (369.62,22.53) .. (372.27,22.53) -- (415.87,22.53) .. controls (418.52,22.53) and (420.67,24.68) .. (420.67,27.33) -- (420.67,41.73) .. controls (420.67,44.38) and (418.52,46.53) .. (415.87,46.53) -- (372.27,46.53) .. controls (369.62,46.53) and (367.47,44.38) .. (367.47,41.73) -- cycle ;
\draw  [color={rgb, 255:red, 74; green, 74; blue, 74 }  ,draw opacity=1 ][fill={rgb, 255:red, 155; green, 155; blue, 155 }  ,fill opacity=1 ][dash pattern={on 0.84pt off 2.51pt}] (363.47,45.53) .. controls (363.47,41.12) and (366.38,37.53) .. (369.97,37.53) .. controls (373.56,37.53) and (376.47,41.12) .. (376.47,45.53) .. controls (376.47,49.95) and (373.56,53.53) .. (369.97,53.53) .. controls (366.38,53.53) and (363.47,49.95) .. (363.47,45.53) -- cycle ;
\draw  [color={rgb, 255:red, 74; green, 74; blue, 74 }  ,draw opacity=1 ] (452.4,25.67) .. controls (452.4,23.02) and (454.55,20.87) .. (457.2,20.87) -- (495.53,20.87) .. controls (498.18,20.87) and (500.33,23.02) .. (500.33,25.67) -- (500.33,40.07) .. controls (500.33,42.72) and (498.18,44.87) .. (495.53,44.87) -- (457.2,44.87) .. controls (454.55,44.87) and (452.4,42.72) .. (452.4,40.07) -- cycle ;
\draw  [fill={rgb, 255:red, 155; green, 155; blue, 155 }  ,fill opacity=1 ][dash pattern={on 0.84pt off 2.51pt}] (448.4,43.87) .. controls (448.4,39.45) and (451.31,35.87) .. (454.9,35.87) .. controls (458.49,35.87) and (461.4,39.45) .. (461.4,43.87) .. controls (461.4,48.28) and (458.49,51.87) .. (454.9,51.87) .. controls (451.31,51.87) and (448.4,48.28) .. (448.4,43.87) -- cycle ;
\draw [color={rgb, 255:red, 74; green, 74; blue, 74 }  ,draw opacity=1 ]   (250,72.75) -- (288.71,72.61) ;
\draw [shift={(291.71,72.6)}, rotate = 179.79] [fill={rgb, 255:red, 74; green, 74; blue, 74 }  ,fill opacity=1 ][line width=0.08]  [draw opacity=0] (8.93,-4.29) -- (0,0) -- (8.93,4.29) -- cycle    ;
\draw [color={rgb, 255:red, 74; green, 74; blue, 74 }  ,draw opacity=1 ]   (251.33,110.5) .. controls (315.71,90.56) and (210.17,85.28) .. (283.33,73.5) ;
\draw [color={rgb, 255:red, 74; green, 74; blue, 74 }  ,draw opacity=1 ]   (340.67,111.5) -- (364,111.5) ;
\draw [shift={(367,111.5)}, rotate = 180] [fill={rgb, 255:red, 74; green, 74; blue, 74 }  ,fill opacity=1 ][line width=0.08]  [draw opacity=0] (8.93,-4.29) -- (0,0) -- (8.93,4.29) -- cycle    ;
\draw [color={rgb, 255:red, 74; green, 74; blue, 74 }  ,draw opacity=1 ]   (179.8,74.2) -- (179.8,219) ;
\draw [color={rgb, 255:red, 74; green, 74; blue, 74 }  ,draw opacity=1 ]   (179.8,74.2) -- (198.8,74.2) ;
\draw [shift={(201.8,74.2)}, rotate = 180] [fill={rgb, 255:red, 74; green, 74; blue, 74 }  ,fill opacity=1 ][line width=0.08]  [draw opacity=0] (8.93,-4.29) -- (0,0) -- (8.93,4.29) -- cycle    ;
\draw [color={rgb, 255:red, 74; green, 74; blue, 74 }  ,draw opacity=1 ]   (180.2,110.6) -- (198,110.94) ;
\draw [shift={(201,111)}, rotate = 181.1] [fill={rgb, 255:red, 74; green, 74; blue, 74 }  ,fill opacity=1 ][line width=0.08]  [draw opacity=0] (8.93,-4.29) -- (0,0) -- (8.93,4.29) -- cycle    ;
\draw [color={rgb, 255:red, 74; green, 74; blue, 74 }  ,draw opacity=1 ]   (180.2,178.6) -- (198,178.94) ;
\draw [shift={(201,179)}, rotate = 181.1] [fill={rgb, 255:red, 74; green, 74; blue, 74 }  ,fill opacity=1 ][line width=0.08]  [draw opacity=0] (8.93,-4.29) -- (0,0) -- (8.93,4.29) -- cycle    ;
\draw [color={rgb, 255:red, 74; green, 74; blue, 74 }  ,draw opacity=1 ]   (179.8,219) -- (197.6,218.66) ;
\draw [shift={(200.6,218.6)}, rotate = 178.9] [fill={rgb, 255:red, 74; green, 74; blue, 74 }  ,fill opacity=1 ][line width=0.08]  [draw opacity=0] (8.93,-4.29) -- (0,0) -- (8.93,4.29) -- cycle    ;
\draw [color={rgb, 255:red, 74; green, 74; blue, 74 }  ,draw opacity=1 ]   (169.98,130.6) -- (179.44,130.56) ;
\draw [color={rgb, 255:red, 74; green, 74; blue, 74 }  ,draw opacity=1 ]   (251.33,110.5) -- (289,110.19) ;
\draw [shift={(292,110.17)}, rotate = 179.53] [fill={rgb, 255:red, 74; green, 74; blue, 74 }  ,fill opacity=1 ][line width=0.08]  [draw opacity=0] (8.93,-4.29) -- (0,0) -- (8.93,4.29) -- cycle    ;
\draw [color={rgb, 255:red, 74; green, 74; blue, 74 }  ,draw opacity=1 ]   (273.56,110.83) -- (272.33,190.17) ;
\draw [color={rgb, 255:red, 74; green, 74; blue, 74 }  ,draw opacity=1 ]   (272.95,150.5) -- (288.42,150.84) ;
\draw [shift={(291.42,150.9)}, rotate = 181.24] [fill={rgb, 255:red, 74; green, 74; blue, 74 }  ,fill opacity=1 ][line width=0.08]  [draw opacity=0] (8.93,-4.29) -- (0,0) -- (8.93,4.29) -- cycle    ;
\draw [color={rgb, 255:red, 74; green, 74; blue, 74 }  ,draw opacity=1 ]   (272.33,190.17) -- (288.01,190.5) ;
\draw [shift={(291.01,190.57)}, rotate = 181.23] [fill={rgb, 255:red, 74; green, 74; blue, 74 }  ,fill opacity=1 ][line width=0.08]  [draw opacity=0] (8.93,-4.29) -- (0,0) -- (8.93,4.29) -- cycle    ;
\draw [color={rgb, 255:red, 74; green, 74; blue, 74 }  ,draw opacity=1 ]   (420,110.17) -- (436.67,110.17) -- (441,110.36) ;
\draw [shift={(444,110.5)}, rotate = 182.6] [fill={rgb, 255:red, 74; green, 74; blue, 74 }  ,fill opacity=1 ][line width=0.08]  [draw opacity=0] (8.93,-4.29) -- (0,0) -- (8.93,4.29) -- cycle    ;
\draw [color={rgb, 255:red, 74; green, 74; blue, 74 }  ,draw opacity=1 ]   (340.33,70.17) -- (357,70.17) -- (364.33,70.17) ;
\draw [shift={(367.33,70.17)}, rotate = 180] [fill={rgb, 255:red, 74; green, 74; blue, 74 }  ,fill opacity=1 ][line width=0.08]  [draw opacity=0] (8.93,-4.29) -- (0,0) -- (8.93,4.29) -- cycle    ;
\draw [color={rgb, 255:red, 74; green, 74; blue, 74 }  ,draw opacity=1 ]   (347,29.67) -- (347,69.83) ;
\draw [color={rgb, 255:red, 74; green, 74; blue, 74 }  ,draw opacity=1 ]   (347,29.67) -- (364.67,29.81) ;
\draw [shift={(367.67,29.83)}, rotate = 180.46] [fill={rgb, 255:red, 74; green, 74; blue, 74 }  ,fill opacity=1 ][line width=0.08]  [draw opacity=0] (8.93,-4.29) -- (0,0) -- (8.93,4.29) -- cycle    ;
\draw [color={rgb, 255:red, 74; green, 74; blue, 74 }  ,draw opacity=1 ]   (340.33,150.5) -- (357,150.5) -- (364.33,150.5) ;
\draw [shift={(367.33,150.5)}, rotate = 180] [fill={rgb, 255:red, 74; green, 74; blue, 74 }  ,fill opacity=1 ][line width=0.08]  [draw opacity=0] (8.93,-4.29) -- (0,0) -- (8.93,4.29) -- cycle    ;
\draw [color={rgb, 255:red, 74; green, 74; blue, 74 }  ,draw opacity=1 ]   (347,190.83) -- (347,150.83) ;
\draw [color={rgb, 255:red, 74; green, 74; blue, 74 }  ,draw opacity=1 ]   (347,190.83) -- (364.67,190.69) ;
\draw [shift={(367.67,190.67)}, rotate = 179.54] [fill={rgb, 255:red, 74; green, 74; blue, 74 }  ,fill opacity=1 ][line width=0.08]  [draw opacity=0] (8.93,-4.29) -- (0,0) -- (8.93,4.29) -- cycle    ;
\draw  [fill={rgb, 255:red, 155; green, 155; blue, 155 }  ,fill opacity=1 ] (408.67,18.5) -- (425.67,18.5) -- (425.67,28.5) -- (408.67,28.5) -- cycle ;
\draw  [fill={rgb, 255:red, 155; green, 155; blue, 155 }  ,fill opacity=1 ] (240.67,204.83) -- (257.67,204.83) -- (257.67,214.83) -- (240.67,214.83) -- cycle ;
\draw  [fill={rgb, 255:red, 155; green, 155; blue, 155 }  ,fill opacity=1 ] (494,17.83) -- (511,17.83) -- (511,27.83) -- (494,27.83) -- cycle ;
\draw  [fill={rgb, 255:red, 155; green, 155; blue, 155 }  ,fill opacity=1 ][dash pattern={on 0.84pt off 2.51pt}] (66.6,204) .. controls (66.6,199.58) and (69.51,196) .. (73.1,196) .. controls (76.69,196) and (79.6,199.58) .. (79.6,204) .. controls (79.6,208.42) and (76.69,212) .. (73.1,212) .. controls (69.51,212) and (66.6,208.42) .. (66.6,204) -- cycle ;
\draw [color={rgb, 255:red, 74; green, 74; blue, 74 }  ,draw opacity=1 ]   (421.67,35.83) -- (438.33,35.83) -- (448.67,35.83) ;
\draw [shift={(451.67,35.83)}, rotate = 180] [fill={rgb, 255:red, 74; green, 74; blue, 74 }  ,fill opacity=1 ][line width=0.08]  [draw opacity=0] (8.93,-4.29) -- (0,0) -- (8.93,4.29) -- cycle    ;
\draw  [color={rgb, 255:red, 74; green, 74; blue, 74 }  ,draw opacity=1 ] (201,138.4) .. controls (201,135.75) and (203.15,133.6) .. (205.8,133.6) -- (246.53,133.6) .. controls (249.18,133.6) and (251.33,135.75) .. (251.33,138.4) -- (251.33,152.8) .. controls (251.33,155.45) and (249.18,157.6) .. (246.53,157.6) -- (205.8,157.6) .. controls (203.15,157.6) and (201,155.45) .. (201,152.8) -- cycle ;
\draw  [color={rgb, 255:red, 74; green, 74; blue, 74 }  ,draw opacity=1 ][fill={rgb, 255:red, 155; green, 155; blue, 155 }  ,fill opacity=1 ][dash pattern={on 0.84pt off 2.51pt}] (197,156.6) .. controls (197,152.18) and (199.91,148.6) .. (203.5,148.6) .. controls (207.09,148.6) and (210,152.18) .. (210,156.6) .. controls (210,161.02) and (207.09,164.6) .. (203.5,164.6) .. controls (199.91,164.6) and (197,161.02) .. (197,156.6) -- cycle ;
\draw [color={rgb, 255:red, 74; green, 74; blue, 74 }  ,draw opacity=1 ]   (180.2,144.6) -- (198,144.94) ;
\draw [shift={(201,145)}, rotate = 181.1] [fill={rgb, 255:red, 74; green, 74; blue, 74 }  ,fill opacity=1 ][line width=0.08]  [draw opacity=0] (8.93,-4.29) -- (0,0) -- (8.93,4.29) -- cycle    ;
\draw (74,184) node  {\includegraphics[width=9pt,height=13.5pt]{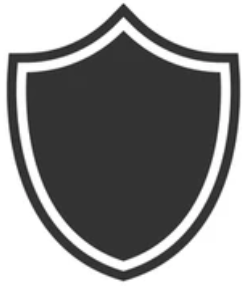}};
\draw (74.1,165.2) node  {\includegraphics[width=9.75pt,height=11.7pt]{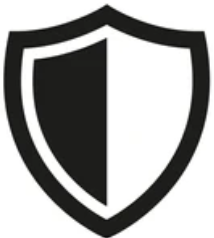}};
\draw (169.7,142) node  {\includegraphics[width=9.75pt,height=11.7pt]{semi.png}};
\draw (249.7,82.8) node  {\includegraphics[width=9.75pt,height=11.7pt]{semi.png}};
\draw (249.7,154.8) node  {\includegraphics[width=9.75pt,height=11.7pt]{semi.png}};
\draw (250.1,188.8) node  {\includegraphics[width=9.75pt,height=11.7pt]{semi.png}};
\draw (251.2,120.4) node  {\includegraphics[width=9pt,height=13.5pt]{malicious.png}};
\draw (250.9,230.4) node  {\includegraphics[width=9.75pt,height=11.7pt]{semi.png}};
\draw (340.4,83.2) node  {\includegraphics[width=9pt,height=13.5pt]{malicious.png}};
\draw (340.8,122.4) node  {\includegraphics[width=9pt,height=13.5pt]{malicious.png}};
\draw (339.6,164) node  {\includegraphics[width=9pt,height=13.5pt]{malicious.png}};
\draw (339.6,203.6) node  {\includegraphics[width=9pt,height=13.5pt]{malicious.png}};
\draw (420.4,82.8) node  {\includegraphics[width=9pt,height=13.5pt]{malicious.png}};
\draw (420.4,121.6) node  {\includegraphics[width=9pt,height=13.5pt]{malicious.png}};
\draw (500.8,120.4) node  {\includegraphics[width=9pt,height=13.5pt]{malicious.png}};
\draw (420.5,202) node  {\includegraphics[width=9.75pt,height=11.7pt]{malicious.png}};
\draw (421.3,45.8) node  {\includegraphics[width=9.75pt,height=11.7pt]{semi.png}};
\draw (420.9,163.2) node  {\includegraphics[width=9.75pt,height=11.7pt]{malicious.png}};
\draw (500.5,43.8) node  {\includegraphics[width=9.75pt,height=11.7pt]{semi.png}};

\draw (208.27,173.93) node [anchor=north west][inner sep=0.75pt]   [align=left] {{\footnotesize Quotient}};

\draw (208.13,105.07) node [anchor=north west][inner sep=0.75pt]   [align=left] {{\footnotesize ABY3}};

\draw (199.4,216) node [anchor=north west][inner sep=0.75pt]   [align=left] {{\footnotesize ParSecureML}};

\draw (203.13,66.33) node [anchor=north west][inner sep=0.75pt]   [align=left] {{\footnotesize SecureNN}};

\draw (294.27,146.53) node [anchor=north west][inner sep=0.75pt]   [align=left] {{\footnotesize Trident}};

\draw (300.33,186) node [anchor=north west][inner sep=0.75pt]   [align=left] {{\footnotesize Flash}};

\draw (301.33,104.2) node [anchor=north west][inner sep=0.75pt]   [align=left] {{\footnotesize Blaze}};

\draw (297.13,67.2) node [anchor=north west][inner sep=0.75pt]   [align=left] {{\footnotesize Falcon}};

\draw (369.6,186.47) node [anchor=north west][inner sep=0.75pt]   [align=left] {{\footnotesize MPCLeague}};

\draw (373.8,145.13) node [anchor=north west][inner sep=0.75pt]   [align=left] {{\footnotesize Tetrad}};

\draw (372.73,103.2) node [anchor=north west][inner sep=0.75pt]   [align=left] {{\footnotesize Swift}};

\draw (446.93,104.47) node [anchor=north west][inner sep=0.75pt]   [align=left] {{\footnotesize Fantastic Four}};

\draw (370.1,66.03) node [anchor=north west][inner sep=0.75pt]   [align=left] {{\footnotesize AdamPriv}};

\draw (369.6,26.6) node [anchor=north west][inner sep=0.75pt]   [align=left] {{\footnotesize CryptGPU}};

\draw (454.8,25.33) node [anchor=north west][inner sep=0.75pt]   [align=left] {{\footnotesize Piranha}};

\draw (200.41,79) node [anchor=north west][inner sep=0.75pt]   [align=left] {{\footnotesize 3}};

\draw (443.54,117.27) node [anchor=north west][inner sep=0.75pt]   [align=left] {{\footnotesize 4}};

\draw (366.61,159) node [anchor=north west][inner sep=0.75pt]   [align=left] {{\footnotesize 4}};

\draw (290.41,159.4) node [anchor=north west][inner sep=0.75pt]   [align=left] {{\footnotesize 4}};

\draw (367.41,198.2) node [anchor=north west][inner sep=0.75pt]   [align=left] {{\footnotesize 4}};

\draw (200.41,115.6) node [anchor=north west][inner sep=0.75pt]   [align=left] {{\footnotesize 3}};

\draw (290.41,79.4) node [anchor=north west][inner sep=0.75pt]   [align=left] {{\footnotesize 3}};

\draw (291.61,118.2) node [anchor=north west][inner sep=0.75pt]   [align=left] {{\footnotesize 3}};

\draw (366.21,117.4) node [anchor=north west][inner sep=0.75pt]   [align=left] {{\footnotesize 3}};

\draw (367.01,79) node [anchor=north west][inner sep=0.75pt]   [align=left] {{\footnotesize 3}};

\draw (366.87,40.53) node [anchor=north west][inner sep=0.75pt]   [align=left] {{\footnotesize 3}};

\draw (290.01,199) node [anchor=north west][inner sep=0.75pt]   [align=left] {{\footnotesize 2}};

\draw (200.01,227) node [anchor=north west][inner sep=0.75pt]   [align=left] {{\footnotesize 2}};

\draw (200.21,186.6) node [anchor=north west][inner sep=0.75pt]   [align=left] {{\footnotesize 2}};

\draw (118.8,128) node [anchor=north west][inner sep=0.75pt]   [align=left] {{\footnotesize SecureML}};

\draw (119.41,138) node [anchor=north west][inner sep=0.75pt]   [align=left] {{\footnotesize 2}};

\draw (448.81,38.87) node [anchor=north west][inner sep=0.75pt]   [align=left] {{\tiny $\geq$ \footnotesize 2}};

\draw (89,179) node [anchor=north west][inner sep=0.75pt]   [align=left] {Malicious};

\draw (70.01,201) node [anchor=north west][inner sep=0.75pt]   [align=left] {{\footnotesize n}};

\draw (89,199) node [anchor=north west][inner sep=0.75pt]   [align=left] {n-party};

\draw (494,20.33) node [anchor=north west][inner sep=0.75pt]   [align=left] {{\tiny GPU}};

\draw (408.67,22) node [anchor=north west][inner sep=0.75pt]   [align=left] {{\tiny GPU}};

\draw (240.67,208.33) node [anchor=north west][inner sep=0.75pt]   [align=left] {{\tiny GPU}};

\draw (210.8,139.73) node [anchor=north west][inner sep=0.75pt]   [align=left] {{\footnotesize ABY2}};

\draw (200.41,150) node [anchor=north west][inner sep=0.75pt]   [align=left] {{\footnotesize 2}};

\draw (87.8,161.4) node [anchor=north west][inner sep=0.75pt]   [align=left] {Semi-honest};

\end{tikzpicture}

}
	\caption{SMPC Taxonomy -- illustrates comparative framework employed in our study, showcasing performance evaluations among different protocols. For example, Falco against SecureNN and ABY3, and SecureNN against SecureML.}
	\label{fig:mpctaxonomy}
\end{figure}

Over the past years, various methods focused on constructing SMPC protocols with different properties and settings. However, as listing all the relevant techniques exceeds the scope of this work we recommend a well-written and friendly introduction to SMPC~\cite{evans2018pragmatic}. Nevertheless, we intend to review the most important SMPC approaches that perform PPML training among multiple parties and show their relationships in~\autoref{fig:mpctaxonomy}. These works are discussed in detailed below. We categorize SMPC techniques based on the number of parties involved in the protocol. 
This categorization can vary from a 2-party protocol, where training is conducted on two non-colluding parties, to a 4-party protocol, where the protocol can be carried out by four parties. The main motivation for increasing the number of parties in the SMPC protocols can be twofold: \begin{inparaenum}[\it (i)] \item aiming for more efficient computations and collaborative tasks with the same amount of corruptions as in lower party variants, or \item  improving security through increased resilience against more potential malicious parties. \end{inparaenum}  However, it is important to note that the specific security guarantees may vary depending on the chosen protocol and configuration.  

\noindent \textbf{Two-party Computation:} 
\label{subsubsec: two party} 
The 2-party computation is a representative of SMPC~\cite{goldreich1998secure}. Many 2-party SMPC based ML models have been developed among which SecureML~\cite{mohassel2017secureml}, proposed by Mohassel \textit{et al.}, was 
the first privacy-preserving protocol for efficient NN training. It is based on SS protocol 
where the data owner distributes private data among two 
servers. 
Compared to previous SotA frameworks in PPML, SecureML is much faster than the protocol implemented in \cite{nikolaenko2013privacy,gascon2016secure}, however, it is an order of magnitude slower than the privacy free counterparts. 
Though SecureML excels in certain aspects of PPML, its limitations in WAN settings (high number of interactions and communication overhead) prompt the exploration of alternative frameworks. QUOTIENT~\cite{agrawal2019quotient} emerges as a compelling choice, showcasing superior speed and accuracy in both WAN and LAN settings. Its core idea lies in ternarizing the network weights into integer values of $\{-1, 0, 1 \}$ upon which the SMPC-aware NN training protocols are developed. 
Compared to SecureML, training with QUOTIENT is~50$\times$ faster in both WAN and LAN setting. By implementing ML features like adaptive gradients and the normalization approach, QUOTIENT reaches the level of accuracy of SecureML relatively fast -- in less than 
two epochs. Furthermore, for the same network in SecureML, QUOTIENT achieves an overall accuracy of~99.38\%, a~6\% improvement over the SecureML (93.4\%).
In order to make a 2-party SMPC technique even more efficient, a different approach is taken by ParSecureML~\cite{chen2020parsecureml}. ParSecureML, leveraging GPU-based parallelization, offers a substantial speedup, presenting a noteworthy alternative for efficient PPML.
It consists of three major parts: \begin{inparaenum}[\it (i)] \item Profiling-guided adaptive GPU utilization, \item double pipeline design for intra-node data transmission between CPU-GPU and \item compressed transmission for inter-node communication between two servers \end{inparaenum}. These three components are integrated in order to enable 2PC on GPUs. Compared to SecureML~\cite{mohassel2017secureml}, the authors show that ParSecureML achieves an average of~32.2$\times$ speedup. Another notable work in the field of 2-party PPML is ABY2.0~\cite{patra2021aby2}. ABY2.0 proposes a mixed-world SMPC protocol between Arithmetic, Boolean and Yao sharing, based on which a NN training protocol was built. For NN training, ABY2.0 has 2.7-3.46$\times$ and 2.4-2.8$\times$ online throughput improvements for LAN and WAN settings respectively.

In a 2-party SMPC protocol, where only two parties are involved, the threat model considered is a weaker semi-honest threat model (see~\autoref{table:smpc comaprison}). This model 
assumes 
parties will follow the protocol 
and this might not always be the case in the real world. If one party is malicious, the system may not work effectively. To address this limitation and take full advantage of the benefits of SMPC, 
aimed 
at facilitating secure collaboration among multiple distrustful parties, researchers are exploring ways to involve more than two parties. The primary goal of adding more parties is twofold: firstly, researchers aim to improve the training process by incorporating a larger dataset — more parties participating implies more data for training. Secondly, this expansion is intended to address a more realistic malicious threat model. This widens the applicability and enhances the security of SMPC in real-world scenarios.



\noindent \textbf{Three-party Computation:} 
\label{subsubsec: three party} 
In 3-party SMPC, 
only one protocol, CryptGPU~\cite{tan2021cryptgpu}, exclusively considers the weaker semi-honest threat model. 
Others 
take into account both threat models. For instance: 
$ABY^{3}$~\cite{mohassel2018aby3} designs techniques for encrypted training of DNNs in the 3-party settings with a single corrupted server. 
This is a mixed protocol framework for ML, which uses SMPC techniques and offers efficient support for fixed point arithmetic computation, improved matrix multiplication (to reduce the amount of communication) and 
an efficient piece-wise polynomial evaluation technique. 
$ABY^{3}$ experiments on both inference and training for LR, LoR and NN models. For training NNs, $ABY^{3}$ is~55,000$\times$ faster than the 2PC solution of SecureML. Another significant contribution in the 3-party SMPC settings is \textit{SecureNN}~\cite{wagh2019securenn}, designed to ensure privacy against one malicious corruption and privacy and correctness against one semi-honest corruption. Prior works to SecureNN use boolean computation and Yao’s Garbled Circuits (GC) to construct functions, such as ReLU, Maxpool and their derivatives. 
This 
results in 
increased communication cost. SecureNN's proposed protocols for Boolean computation require much lesser communication overhead than the cost of converting to a Yao encoding and executing a GC. Thanks to 
protocol efficiency, 
SecureNN is the first work to 
privately train a Convolutional Neural Network (CNN)~\cite{albawi2017understanding} 
with an accuracy of higher than~99\% on MNIST. Compared to SecureML, SecureNN is~79$\times$ faster in the LAN (latency~0.22 ms, bandwidth~625 MB/s) setting, and~553$\times$ faster in the WAN setting (latency~58 ms, bandwidth~40 MB/s). In the 3-party LAN setting, SecureNN outperforms SecureML in training time by~7$\times$. Moving forward, Patra and Suresh proposes a PPML framework named \textit{BLAZE}~\cite{patra2020blaze} in the 3-party setting, tolerating one malicious corruption over a ring ($Z_{2^\ell}$). 
It consists of a data-independent preprocessing phase, 
used to perform a relatively expensive computation, and a fast input-dependent online phase. 
The authors benchmark the performance of the framework over $ABY^3$ for training LR and LoR models. 
Over a dataset with a feature size of 784 and batch size of 128, training in Blaze gains about 4$\times$ more throughput in the preprocessing phase for both LR and LoR. For the online phase, Blaze gains 145.35$\times$ and 31.89$\times$ throughput for LR and LoR respectively compared to $ABY^3$. 
Shifting focus to \textit{Falcon}~\cite{wagh2021falcon}, which is an end-to-end 3-party protocol for efficient private training and the inference of large ML models. 
By combining techniques from SecureNN and $ABY^{3}$, Falcon constructs improved protocols for various private ML operations, such as convolutions, matrix multiplications, private comparison, ReLU, and its derivative, division, and batch normalization. Compared to other 
private training frameworks, Falcon is about~4.4$\times$ faster than $ABY^{3}$ and~6$\times$ faster than SecureNN. Furthermore, Falcon achieves~2$\times$ to 60$\times$ less communication overhead than $ABY^{3}$ and SecureNN. Similarly, \textit{SWIFT}~\cite{koti2021swift} relies on an efficient, malicious-secure 3PC framework, that works over rings ($\mathbb{Z}_{2^\ell}$ and $\mathbb{Z}_{2^1}$). SWIFT provides Guaranteed Output Delivery (GOD) in the honest majority setting. The protocols work in the preprocessing (offline-online) model. One highlight contribution is the dot product protocol that achieves communication cost independent of the input vector sizes. 
The authors benchmark their method using LoR (for training and inference), LeNet~\cite{lecun2015lenet} and VGG16~\cite{simonyan2014very} NNs (for inference). Authors compare LoR model's training in a 3-party setup with BLAZE, finding similar performance and improved security.

Adam in Private~\cite{attrapadung2022adam} addresses specific tasks within a 3-party setting and specializes in tasks such as integer division, exponentiation, inversion, and square root extraction within the context of DNNs.
To demonstrate the proposed protocol's scalability, 
the authors perform measurements on DNNs architectures, such as a 3-layer fully-connected network introduced in SecureML, AlexNet and VGG16 on two datasets --MNIST and CIFAR10. The experiments are carried out in both LAN and WAN and the results are compared to Falcon. The implementation of the Adam optimization algorithm allows the framework to converge faster and requires fewer epochs compared to Falcon. In terms of training time, Adam in Private is~3.2$\times$ to~6.7$\times$ faster than Falcon for the 3-layer NN, about~12$\times$ to 14$\times$ faster for AlexNet and~46$\times$ to 48$\times$ faster for VGG16.

While the above works demonstrate a CPU-only implementation, there are a few works that explore GPU assisted computation within the standard 3-party setting such as \textit{CryptGPU}~\cite{tan2021cryptgpu} and \textit{Piranha}~\cite{watson2022piranha}. CryptGPU operates, where all inputs are secretly shared among three non-colluding servers executing the protocol. It is built on top of PyTorch~\cite{paszke2019pytorch} and Crypten~\cite{knott2021crypten}. 
Experiments show, that the GPU-based protocol can have a~150$\times$ speed up over the CPU-based protocol for a convolution operation and up to~10$\times$ the speed up for non-linear operations. 
The increased efficiency of 
GPU implementations, 
enables the framework 
to privately train big networks such as LeNet, AlexNet~\cite{krizhevsky2012imagenet}, VGG16 over MNIST, CIFAR-10 and ImageNet datasets. Compared to Falcon, CryptGPU achieves~6$\times$ to~36$\times$ improvement for private training.
Though showing 
great progress in terms of GPU adoption, CryptGPU is tailored for a specific SMPC protocol and can only demonstrate full end-to-end training 
on simple networks such as AlexNet. To improve upon this, \textit{Piranha}~\cite{watson2022piranha} proposed a more general GPU-based framework that can be used to implement other SMPC protocols. Piranha consists of three layers: 
On the device layer, Piranha implements a data abstraction that manages vectorized GPU data and integer GPU kernels to support acceleration for integer-based computation needed for SMPC protocols. Piranha's protocol layer uses the device layer to implement functionality for different SMPC protocols, namely SecureML (2-party)~\cite{mohassel2017secureml}, Falcon (3-party)~\cite{wagh2021falcon}, and FantasticFour (4-party)~\cite{dalskov2021fantastic}. In the application protocol, Piranha provides typical privacy-preserving layers for NNs such as the linear and convolution layers, pooling operations, the ReLU activation function, and layer normalization. The SMPC protocols implemented with Piranha exhibit 16-48$\times$ training time speed up compared to their native CPU implementations. Piranha is also the first PPML framework capable of training a big NN such as VGG16 (with over 100 million parameters) over the CIFAR10 dataset in a short amount of time (less than a day and a half).

\noindent \textbf{Four-party Computation:} 
\label{subsubsec: 4pc} 
As we move forward, the ongoing evolution of 3-party SMPC protocols not only enhances efficiency, security, and applicability in collaborative computations but also lays the foundation for exploring more complex scenarios. This progress is evident in recent developments within 4-party SMPC protocols, where innovative solutions address diverse challenges in secure computation. For example:
\textit{Trident}~\cite{chaudhari2019trident} proposes a framework in the setting of 4-parties with at most one corrupt participant over the ring  $\mathbb{Z}_{2^\ell}$. The sharing protocols follow the offline-online phase paradigm and are used to construct a mixed-world framework between the binary, arithmetic and GC worlds. 
Compared to $ABY^3$~\cite{mohassel2018aby3} (for LR), training with Trident is~4.88$\times$ to~251.84$\times$ faster in LAN and~2$\times$ to~2.83$\times$ in WAN. For the LoR model, Trident's improvements range from~5.95$\times$ to~67.88$\times$ in LAN and~2.71$\times$ to~2.96$\times$ in WAN. Training the NNs, Trident is from~3.56$\times$ to~62$\times$ faster in LAN, and~2.97$\times$ to~3.56$\times$ faster in WAN. 
Building on the foundation laid by Trident, \textit{MPCLeague}~\cite{suresh2021mpcleague} operates within a 4PC setting over the ring $\mathbb{Z}_{2^\ell}$, ensuring support for an honest majority with at most one corrupted party. The framework combines arithmetic, boolean and garbled worlds with efficient end-to-end conversions between them. 
The efficiency of MPCLeague in private NN training is benchmarked using networks such as LeNet and VGG16. Compared to Trident~\cite{chaudhari2019trident} based on runtime, communication, and cost, MPCLeague offers better performance overall. 

In the same year, Koti \textit{et al.} introduced \textit{Tetrad}~\cite{koti2021tetrad}, a 4-party setting designed to tolerate at most one active corruption over the ring $\mathbb{Z}_{2^\ell}$. Tetrad proposes a mixed SMPC protocol that supports robustness and fairness. 
Similar to MPCLeague, the authors of Tetrad compared their results with Trident 
considering training time, communication and monetary cost, as this 
captures the effect of the total runtime 
and communication of the parties. 
Experimental results show that Tetrad is~3-4$\times$ faster than Trident 
and achieves approximately 30\% better results 
in terms of monetary cost. To obtain security against malicious parties, Dalskov \textit{et al,} proposed Fantastic Four~\cite{dalskov2021fantastic}, an actively secure 4-party protocol for corruption over a ring $\mathbb{Z}_{2^k}$.
The protocol tolerates one active corruption and satisfies security with abort, however, it also provides GOD with some extensions.
The protocol guarantees to identify a semi-corrupt pair and remove one party in the pair, then proceeds to an actively secure 3-party protocol with abort. If the 3PC protocol succeeds, then the output is produced, and the computation is finished. If the 3PC protocol also aborts, then the protocol knows that it removed an honest party and will abort 
the remaining party in the corrupt pair. It then proceeds to a passively secure 2-party protocol. 
The protocol is applied 
for training fully-connected NNs with~1-3 dense layers and a LoR model on the MNIST dataset. For the LoR model, Fantastic Four achieves~172.05$\times$ faster training time than SWIFT in 
4PC setting, and~84.24$\times$ in 
3PC setting.

To summarize, the choice among 2-party, 3-party and 4-party SMPC depends on the specific requirements and characteristics of the collaboration at hand. The lower party like the 2-party SMPC may be favored for simplicity and lower computational overhead, whereas the more intricate 4-party SMPC offers advantages such as enhanced collaboration, security, and flexibility in situations involving more than two parties. Adding more parties may offer better security and efficiency, 
though it necessitates more servers, 
more communication overhead, more efforts to coordinate and increased complexities in the algorithms.

\section{Evaluation}
\label{section:evaluation}

\begin{center}
    \begin{tcolorbox}[width=0.50\textwidth,title={\textbf{Reproducible Research: An Unfulfilled Dream}}, colbacktitle=gray]
In this section, we evaluated both the performance and security of several PPML works. 
At first, we intended to benchmark and compare~23 impactful PPML techniques based on HE and SMPC. As a result, we made an effort to reproduce the results for all these works. However, this 
proved challenging because most of the 
implementations did \textit{not} provide open-source 
code. We hence tried to reimplement some of the schemes following the instructions given in the respective papers. 

  However, implementation of HE and SMPC techniques proved to be 
monumental tasks and we were unable to reproduce accurate results, because of
difficulties, such as time constraints, hardware limitations, and library issues.
   Although 
a few works have OSIs, they lack proper documentation, making it challenging to use those works. 
Also, comparing works that are 
exclusively designed for either CPU or GPU can be challenging because it restricts cross-platform evaluation. Additionally, given our focus on SMPC-based techniques, variations in the number of parties involved, such as two-party or three-party computation, can result in different performance characteristics, making comparisons more complex. While the ideal scenario would involve a universal framework evaluation under various settings, this would demand substantial engineering effort and computational resources, which are currently beyond our capacity. Consequently, we were only able to reproduce the results of~4 PPML techniques, the details of which are given in~\autoref{tab:trainingResultsSMPCFrameworks}. Despite the fact that solutions are 
comparable in terms of metrics 
-i.e. computational and communication costs
accuracy and memory usage- unfortunately, not all solutions provide 
said metrics (see~\autoref{tab:trainingResultsSMPCFrameworks}), 
thus rendering 
direct comparisons difficult. 
Despite the mentioned difficulties, we hope to give some valuable insights to researchers who wish to further expand the area by covering the results of SotA in the field and 
mentioning some of their existing limitations. 
    \end{tcolorbox}
\end{center}

    \subsection{Experiments}
\label{subsec:experiments}

\noindent\textbf{HE Protocols:}
\label{subsubsec:heprotocols-experiments}
 In this section, we overview results from available HE frameworks. We also 
cover aspects relating to their performance and their availability, while equally focusing on 
the specifications used by the researchers during testing. 

\noindent\textit{Testing environments.}
As covered in the preliminaries section, 
HE is computationally expensive to properly implement and test. 
This fact can be corroborated by the implementations 
covered as in the majority used high-performance server processors, such as the Intel Xeon series and considerable amounts of RAM (i.e.\ ranging from~48 to~256GB). 
An exception to the high requirements is the implementation made by Mihara \textit{et al.}~\cite{mihara2020neural} 
requiring only consumer-level hardware, such as the Intel Core i7 and up to 32GB of RAM. 
All of the implementations run a Unix-based operating system, such as Linux, through Ubuntu or ArchLinux. 
As can be seen in ~\autoref{table:results-of-he}, the high computational requirements of HE create an entry barrier making it impossible for most users to use HE technology outside commercial companies and academia. However, HE can still be used in cloud services. In this case, most users can use the services, while keeping their personal information private.

\begin{center}
    \begin{tcolorbox}[width=0.50\textwidth,title={\textbf{Reproducible Research: Myth Buster \#1}}, colbacktitle=gray]
	When overviewing the availability of HE frameworks for result replicability and further testing, it can be noted that most frameworks do \textit{not} release an OSI of their code. 
Hence, 
the majority of frameworks extensively cover their implementation 
\textit{(a)} by providing 
libraries 
used for the implementation (see~\autoref{fig:hetaxonomy} and~\autoref{subsec:analysis-he}) 
and \textit{(b)} by defining the functions and model parameters for others to see. Most of the works provide a pseudocode and algorithms for their implementation, however, 
none of the surveyed HE works 
provided an openly available code.
As mentioned 
before, it is possible to recreate the core aspects of private implementations with the tools provided in the 
papers. However, the main drawback in rewriting the code from scratch is that the resulting code 
might produce large deviations from the original results, 
due to the naturally occurring differences in algorithms.
 Another aspect to be taken into account is amount of time needed to recreate the code as well as implement HE into ML techniques.
 \textit{As such the lack of OSIs creates possible gaps in reaffirming results and advancing science.}
    \end{tcolorbox}
\end{center}

\noindent\textit{Performance.} 
An important topic when covering ML techniques is 
overall performance both in terms of accuracy and training time. Hence, it is important to overview the performance of HE techniques 
introduced into ML and compare their results. 
Following the results listed in~\autoref{table:results-of-he}, one can notice that 
the accuracy of almost all the frameworks is more than~90\%. 
There are two exceptions regarding accuracy: \textit{(a)} the hybrid technique POSEIDON~\cite{sav2020poseidon} and \textit{(b)}
the implementation proposed by Bonte and Vercauteren~\cite{bonte2018privacy}. For POSEIDON, 
accuracy is just short of 90\% on the MNIST dataset (\textit{Accuracy: 89.9\%}), however POSEIDON has been tested on multiple other datasets, including the Breast Cancer Wisconsin (BCW) dataset (\textit{Accuracy: 96.9\%}), the Epileptic Seizure Recognition (ESR) dataset (\textit{Accuracy: 90.4\%}) and the default of credit card clients (CREDIT) dataset (\textit{Accuracy: 80.2\%})~\cite{sav2020poseidon}. 
Bonte and Vercauteren~\cite{bonte2018privacy} tested their implementation on two datasets: \textit{(i)} the iDASH genomic dataset~\cite{tang_wang_wang_jiang_2018}, which 
showed lower accuracy after having trained with different parameters (\textit{Average accuracy: 64.37\%}) and \textit{(ii)} a private financial dataset, 
whose accuracy was higher on average after having trained with different parameters (\textit{Average accuracy:~92.38\%}). 
Other implementations, such as CryptoDL~\cite{hesamifard2018privacy}
show high levels of accuracy -- reaching up to~99\%. This, in turn, shows that HE 
techniques can provide the same level of accuracy as their plaintext counterparts, when compared for the same dataset and NN architecture. 

However, the main concern about HE is the high computational complexity as mentioned in prior sections. That is why when overviewing the training times of all of the frameworks we can notice, 
most implementations can take multiple hours or even days to fully train on different-sized datasets. When comparing HE techniques, which used the MNIST dataset for training, we observe that Nandakumar~\textit{et al.}~\cite{nandakumar2019towards} takes the longest to train (approximately~60K-846K hours) followed by Sphinx~\cite{tian2022sphinx} (approximately~1560 hours). The extremely long training time for Nandakumar~\textit{et al.} is theoretically calculated 
by combining the reported training time in hours for a single mini-batch (reported to 
vary between 0.667h - 9.4h), multiply 
by the total amount of mini-batches (reported to be 1,800) and then multiplying 
by the amount of epochs 
required to train a plaintext model of the same size (reported to be 50)~\cite{nandakumar2019towards}. Similarly, training time for Sphinx~\cite{tian2022sphinx} is calculated by taking the training time for a single batch (reported to be~0.108h) multiplying it by the amount of batches (calculated by dividing the total amount of training samples by the batch size~$\frac{60000}{500}=120$) and then multiplying 
by the amount of epochs (reported to be~120 for MNIST). However, it is important to note that the model used in Nandakumar~\textit{et al.}~\cite{nandakumar2019towards} contains only~3 fully connected layers with a sigmoid activation function, while Sphinx contains~2 convolution,~2 average pooling and~2 fully connected layers with ReLU activation functions. Nandakumar~\textit{et al.} set a solid baseline for further research into FHE applicability in 
PPML environments. Since 
this was one of the first attempts to train an NN in a privacy-preserving way using HE, 
the training time is the longest 
of the covered implementations. 
The performance of newer implementations, such as POSEIDON, CryptoDL and Glyph, increased together with the knowledge 
on PPML techniques. Training times in these implementations are more manageable. 
The shortest training time is proposed in POSEIDON~\cite{sav2020poseidon} (approximately~1.47h) and  
involves 
using FL and HE. 
However, even though the training time of HE techniques has 
dropped considerably in the past years, they 
are multiple times slower to train, when compared to plaintext counterparts. 
This 
increase in 
training time may be the leading issue impeding consumers and suppliers from moving away from PPML techniques, even though HE techniques provide SotA privacy for their data.

\noindent\textbf{SMPC Protocols:}
\label{subsubsec:smpcprotocols}
We evaluate SMPC frameworks 
with an open-sourced code on the MNIST dataset over popular network architectures in PPML domain. The results are reported in~\autoref{tab:trainingResultsSMPCFrameworks}.

\noindent\textit{Testing Environment.} 
To reproduce the results for SMPC, first we created four Virtual Machines (VM) on a locally deployed cloud. 
The experiments are conducted using 
four VMs equipped with Ubuntu~18.04.2: LTS, Intel(R) Xeon(R) Gold 6130 CPU @ 2.10GHz,~16 CPUs,~64 GB RAM and~256 GB Disk. As our VMs are not equipped with dedicated graphic cards, 
we cannot experiment with works such as ParSecureML and CryptGPU.

\noindent\textit{Performance.} 
Below we will discuss the results as well as the issues we faced while implementing the surveyed papers. 

\begin{center}
	\begin{tcolorbox}[width=0.50\textwidth,title={\textbf{Reproducible Research: Myth Buster \#2}}, colbacktitle=gray]
		Our initial goal was to replicate the results for all 
		SMPC works that we surveyed; 
		however, we were able to replicate the results of a few works 
		for local hosts and did not conduct any experiments for LAN or WAN. The reason for this is that we ran into a number of problems, while attempting to reproduce the results for these works such as 
		lack of availability; only 8 of the 15 works provide OSI. 
		Also, some works, that provide OSI, lack proper documentation for implementing the code. 
	\end{tcolorbox}
\end{center}

\textit{SecureML}: Although SecureML did not provide an OSI of their work, we did locate 
one GitHub repository~\cite{Shreya:Lib} 
referring to the implementation for SecureML. 
First and foremost, the author made a commendable contribution 
by implementing SecureML and provide it as open-source is commendable; however, building SecureML from this repository is not simple. 

Second, this repository only contains code for secure two-party LR; there is no implementation for LoR and NNs, while in the paper, the author proposes a secure two-party approach to train LR, LoR and NNs. Finally, only one performance metric 
(accuracy) is considered, while 
other performance metrics, such as communication cost, and time are ignored. For this survey, we reproduce the SecureML results only for the local host, 
receiving an accuracy of 92.98\% for 100 epochs. 

\textit{SecureNN}: The authors provide an OSI of SecureNN that includes ML training and inference. 
We only performed experiments for training where three parties locally trained the LeNet model. 
The results reproduced only consist of training time and communication cost (see~\autoref{tab:trainingResultsSMPCFrameworks}). The training time for LeNet (1 epoch) is~136.533sec, and the total communication cost is~21058.3MB, which is quite efficient in terms of time and cost.


\textit{Falcon}: Falcon produces fast training, however, the code provided does not produce 
accuracy information after training and is limited 
to training time and communication. 
The communication overhead for LeNet may also be erroneous, as running the code shows that 
first-party communication 
is 3346.56MB, 
hence 
bigger than the total communication 
attributed 
to the whole protocol as shown in~\autoref{tab:trainingResultsSMPCFrameworks} (
only 1800 Mb). Based on this information, the communication cost for LeNet can be approximated to 
around 10,000MB for 3 parties. We submitted this issue, however, the repository seems to 
no longer be under active maintenance.



\textit{AriaNN}: We reproduced the results for AriaNN (Network 1 and 2), where the accuracy was 97.96\% and 97.92\%, respectively as in~\autoref{tab:trainingResultsSMPCFrameworks} compared to 98.0\% and 98.3\% from~\cite{ryffel2020ariann}. However, for LeNet, we only got 89.36\% compared to 99.2\% reported in~\cite{ryffel2020ariann}. We also got longer training times, most likely due to our less capable machines. On the other hand, we calculated and reported the communication for private 
training of each epoch, which was not reported in~\cite{ryffel2020ariann}.

\subsection{Analysis of HE protocols}
\label{subsec:analysis-he}

We analyze and summarize various aspects of HE protocols introduced in~\autoref{subsec:HEProtoc}. 


\begin{center}
	\begin{tcolorbox}[width=0.50\textwidth,title={\textbf{Reproducible Research: Myth Buster \#3}}, colbacktitle=gray]
		\textit{Quotient, Blaze, SWIFT, Adam in Private, Trident, MPCLeague, Tetrad}, do not provide OSIs, and others (\textit{Fantastic Four, $ABY^{3}$}) are extremely difficult to implement. Also, the authors of \textit{ParSecureML and CryptGPU} ran their experiments on high-end accelerators such as NVIDIA Tensor Cores and NVIDIA Tesla~V100 GPUs. As our VMs are limited to only CPUs, we were unable to run and reproduce the results of these papers.
	\end{tcolorbox}
\end{center}

\noindent \textit{\textbf{Security:}} 
One of the most important factors to take into account, when analyzing and assessing various secure computation protocols. In this context, almost all of the HE protocols show security strength of approximately more than~80 bits. This is the minimum required strength for non-federal government information, but does not provide enough security as documented in the NIST Special Publication 800-57~\cite{barker2006recommendation}. As mentioned in that report, the security strength defines “a number associated with the amount of work (i.e., the number of operations) that is required to break a cryptographic algorithm or system”. When looking over the works analyzed, the protocol proposed by Bonte and Vercauteran~\cite{bonte2018privacy} estimates, that the security strength is equal to 78 bits, 
namely 
below the previously mentioned minimum threshold. Other protocols such as the one proposed by Nandakumar \textit{et al.}~\cite{nandakumar2019towards} and Glyph~\cite{lou2020glyph} note a security strength of 80 bits, while most others~\cite{hesamifard2018privacy,sav2020poseidon,tian2022sphinx} 
approximate a security strength, 
that is comparable to AES-128. 
The largest documented security strength 
is produced from the PrivFT~\cite{al2020privft} protocol, which estimates a strength of 140 bits.

Unlike the other HE implementations, Sphinx~\cite{tian2022sphinx} 
makes use of DP to increase its resilience to data reconstruction attacks. The authors test their implementations resilience to a gradient matching attack, which aims to recover input images and their labels from the intermediate gradients~\cite{tian2022sphinx}. From their testing, they show that Sphinx outperforms an equivalent DP-only defence mechanism irrespective to the chosen privacy parameters.

Since HE is 
a relatively new way of implementing PPML,
its security 
against ML attacks, such as poisoning attacks, model inversion attacks or reconstruction attacks, has not yet been tested or documented to the best of our knowledge (aside from the gradient matching attack covered in Sphinx~\cite{tian2022sphinx}). 
However, as 
mentioned by researchers, one of the main issues for HE in PPML is finding a suitable compromise between 
efficiency, accuracy and privacy provided by 
HE algorithms~\cite{xu2021privacy}, that would enable the 
HE-based protocol to 
provide effective privacy guarantees without compromising the 
computation times 
or accuracy of the ML algorithm.

\noindent \textit{\textbf{Encryption Schemes and HE libraries:}} 
Different HE schemes allow for tailored use which increases the accuracy and efficiency of the selected use cases. Because of this each analysed paper made use of the common HE schemes discussed in~\autoref{sec: preliminaries}. BGV is used by Nandakumar \textit{et al.}~\cite{nandakumar2019towards} and CryptoDL~\cite{hesamifard2018privacy}, while Bonte and Vercauteran~\cite{bonte2018privacy} use the FV scheme. 
PrivFT~\cite{al2020privft}, Sphinx~\cite{tian2022sphinx}, Mihara \textit{et al.}~\cite{mihara2020neural} and POSEIDON~\cite{sav2020poseidon} use CKKS, while Glyph~\cite{lou2020glyph} uses the TFHE encryption scheme. Each paper made use of HE libraries to implement operations on encrypted data and developed their solutions through the use of them. 

In terms of HE libraries, the most commonly used is Microsoft SEAL~\cite{MS:Lib:SEAL} 
It supports 
BFV and CKKS schemes. As can be seen in~\autoref{fig:hetaxonomy}, SEAL is used in~\cite{al2020privft, mihara2020neural}. HElib~\cite{HELib} 
another popular and early FHE library represented in this work, supports the BGV and CKKS schemes. HElib is used in paper~\cite{hesamifard2018privacy, nandakumar2019towards, lou2020glyph}. Both HElib and SEAL are widely used for binary plaintext spaces. They construct binary circuits to compute the desired functions over encrypted data. They do, however, provide the option 
of a larger plaintext space in situations, where the functions can be evaluated more efficiently, when represented by a modular arithmetic circuit~\cite{aguilar2018comparison}. The FV-NFlib~\cite{FV-NFLlib} 
library only supports BFV scheme. According to~\cite{korchi2019practical}, FV-NFLib is faster than SEAL library. 
However, FV-NFLib does not support high-level circuits 
(only up to~6 levels). 
Therefore, the use of 
FV-NFLib  is recommended for small circuits, while 
SEAL 
for larger ones. 
Lattigo~\cite{Lattigo:Lib} 
also supports 
BFV and CKKS schemes and their respective multiparty version. It is written in 
Go language and performs similarly to cutting-edge C++ libraries (HElib, FV-NFLib, SEAL). Other recent and promising FHE libraries include OpenFHE~\cite{OpenFHE:Lib} 
and the Zama.ai libraries~\cite{Zama:Lib}. 
OpenFHE is a C++ library and supports most modern FHE schemes, such as BGV, BFV, CKKS, FHEW, and TFHE, as well as multiparty extensions for BGV, BFV and CKKS. One of the libraries' major focuses is usability. This is achieved by streamlining the parameter selection process and using the same common API for functions in different schemes. Zama.ai hosts the libraries' TFHE-rs~\cite{Zama:Lib:Concrete} 
and Concrete ML~\cite{Zama:Lib:Concrete:ML}. 
Both of the Zama libraries focus on allowing non-experts of cryptography to implement FHE into their applications.

\noindent \textit{\textbf{Computational Complexity:}} 
The biggest challenge for 
HE 
is the high computational complexity required to train NNs. This 
has held HE back 
from being included in 
modern-day use. With current-day improvements to all HE algorithms and 
hardware development, the required computations and the time it takes to compute them have vastly improved 
compared to older HE implementations. However, despite 
said improvements the computation time is still comparatively high and requires high-end machines to implement HE algorithms. This 
is noted in the results provided 
by researchers, who propose HE implementations: 
the training time 
expands from multiple hours~\cite{sav2020poseidon, nandakumar2019towards, hesamifard2018privacy, nguyen2023split, khan2023more} to multiple days~\cite{lou2020glyph,al2020privft} 
for comparatively simple datasets, such as MNIST 
requiring little-to-none preprocessing. Current-day plaintext implementations of NNs using MNIST can train hundreds of epochs on complex NNs in the span of a couple of minutes~\cite{9017036}, while most HE solutions require 
notably more time for $\leq$10 epochs. As noted by researchers, the training time vastly increases, when training on more complex datasets, such as CIFAR-10 or CIFAR-100~\cite{sav2020poseidon,hesamifard2018privacy}.

\noindent\textit{\textbf{Adversarial Model:}} 
The adversarial model for 
HE-based PPML works 
were defined as having a semi-honest adversary. In this case, the adversary 
can only passively listen and gather information from an available source, such as the dataset. This model is plausible for various real-world applications, where a client would want to store and classify data in a Cloud Service Providers (CSP). However, new discoveries could be 
envisaged regarding security of HE protocols, if assumption is changed to malicious adversaries, which have more tools for breaking security of the algorithms.

\noindent \textit{\textbf{Scheme Applicability:}}
Given the computational complexity of HE, it is important to identify use cases 
where HE can effectively operate within its limitations. 
As such, HE provides the needed capabilities for directly implementing privacy-preserving MLaaS. HE requires high computational capabilities, which can be provided by a CSP. 
A user would be able to design and train their own 
model, despite their 
computational capabilities, by hosting a sever on a CSP and providing the needed training HE data. 
The main decision for the user is choosing the correct HE scheme for 
each task. Both BGV and FV are malleable to a variety of applications, but require quantization from floating point to integers. This 
can cause 
loss of precision and reduced accuracy. 
As a result, these schemes are best suited 
for use cases where precise values are 
unnecessary (
i.e.\ image classification). On the other hand, CKKS allows operations to be accurately performed on floating point values and can be more suited for implementing non-linear activation functions through polynomial approximation. This allows CKKS to 
be used in precise tasks involving one-dimensional medical data as well as 
in object detection. Lastly, TFHE benefits from having fast bootstrapping operation and allows 
simplified implementations of non-linear functions as they would not require polynomial approximation. Through TFHE users would be able to construct more complex models, than the other schemes, but can face issues of scalability because of look-up table storage costs. This leads TFHE to be more suited for natural language processing tasks and 
complex data structures.

\subsection{Analysis of SMPC protocols}
\label{subsec:analysis-mpc}
While considerable progress has been made regarding the efficiency of SMPC protocols, some of the current approaches remain computationally expensive and do not scale well with the types of NNs typically used in modern ML systems. Another significant problem is the requirement for continuous data transfer between parties and for their continuous online availability. 

In this section, we provide a more in-depth comparison and systematization of the SMPC protocols 
summarized in~\autoref{subsec: secMPC}. We discuss their strengths and weaknesses 
with the aim of outlining current challenges that need to be addressed. We look at the privacy and security guarantees provided by these protocols as well as the data type supported and the evaluation parameters. 
We also check whether the given SMPC protocols 
provide an OSI 
(see~\autoref{tab:smpcimplementationcomaprison}).

\noindent \textit{\textbf{GPU Utilization:}} 
GPUs 
are considered one of the most significant foundations for the resurgence of ML, because their parallel architecture is well-suited to dense matrix operations. 
Consequently, ML frameworks, such as TensorFlow, PyTorch, and Caffe allow GPU acceleration. As GPUs played an important role in the success of modern ML techniques, they 
also became essential for 
scalable PPML. However, in the literature, most of the works on PPML are CPU-based and only two works consider the GPU-based research for PPML~\cite{tan2021cryptgpu, chen2020parsecureml}. One thing to remember is that while choosing and designing PPML protocols for the GPU, one must carefully calibrate them for the architecture. Protocols like Yao's GC are less well suited for taking advantage of GPU parallelism compared to an SS-based protocol. Similarly, protocols that require extensive finite field arithmetic will incur more overhead on the GPU compared to the protocols that only rely on arithmetic modulo of a power of~2.

\noindent \textit{\textbf{Security:}} 
Currently 
the fastest SMPC protocols only provide security with abort. 
This means 
that a malicious service provider will cause the computation to abort without any output 
thus rendering the later appearance of the input provider irrelevant. 
Some SMPC techniques 
provide fairness. 
Protocols 
providing security with abort or fairness will not suffice as in both 
cases an adversary can 
cause the protocol to abort, 
thus not producing the desired output for the user. This leads to denial of service and heavy economic losses for the service provider. Therefore, some SotA SMPC approaches, as 
described in~\autoref{tab:smpcimplementationcomaprison} ensure robustness, guaranteeing that the correct output is produced, no matter how the adversary behaves.

\noindent \textit{\textbf{Secure Machine Learning (SML):}} 
SML refers to preventing leakage of user information 
by protecting the process of ML. As can be seen in~\autoref{tab:smpcimplementationcomaprison}, different ways are used to achieve SML. 
In paper~\cite{mohassel2017secureml, agrawal2019quotient, chen2020parsecureml, ryffel2020ariann}, the authors used two-party computation, in papers~\cite{wagh2019securenn, mohassel2018aby3, patra2020blaze, wagh2021falcon, koti2021swift, tan2021cryptgpu, attrapadung2022adam}, the authors used three-party computation, while in papers~\cite{chaudhari2019trident, dalskov2021fantastic, suresh2021mpcleague, koti2021tetrad} the authors used four-party computation. In~\autoref{tab:smpcimplementationcomaprison}, we also consider the input/output or model privacy that the above works aim 
to protect. As input privacy is the key feature of PPML, 
all the above works provide input privacy. However, only few works incorporate 
output and model privacy (broader and active area of research).

\noindent \textit{\textbf{Adversarial Model (AM):}} 
AM defines the threats and adversary capabilities on a cryptographic protocol. It can be classified either based on the adversarial behavior or on the number of corruptions~\cite{lindell2020secure}. Based on their behavior, the adversaries are categorized into semi-honest and malicious. In a semi-honest setting, the adversary follows the protocol, but tries to glean additional information from the message. Most of the literature 
usually assumes a semi-honest adversary as 
mentioned in~\autoref{table:smpc comaprison}. Said 
adversary 
is limited in its offensive capabilities. 
This 
type of adversary model lowers the performance requirements. In malicious behavior, the adversary arbitrarily deviates from the protocol. 
This requires the adversary to either follow the protocol or do something completely different. The second one is based on the number of corruptions, which can be further classified into two categories: honest and dishonest majority. If $N$ parties are taking part in SMPC, then in honest majority at most $(\frac{N}{2} - 1)$ are allowed to be corrupt, ensuring that the number of honest parties is in the majority. Adversely, in dishonest majority parties are allowed to be corrupt only as high as $N-1$. 
\noindent \textit{\textbf{Data Types:}} 
Any computable function can be securely evaluated in SMPC using two types of secret sharing: Additive Secret Sharing (ASS)~\cite{blakley1979safeguarding} and Boolean Secret Sharing (BSS)~\cite{wang2007two, pullonen2015combining}. In ASS, the data is additively shared between the parties. For example, 
$D$ is the original dataset, which is then shared 
to two parties: 
one party has the share $D1$, while another party has the share $D2$. $D1$ and $D2$ can reconstruct $D$ by adding their data together. Therefore, as long as they do not collude, the original dataset $D$ is kept private. ASS uses an arithmetic circuit and supports efficient calculations over integers and floating-point, 
that do not involve comparisons. In these protocols, additions are cheap, whereas multiplications are expensive.  
Alternatively, in BSS the data is XORed shared between the parties (bit by bit). For BSS, the SMPC uses boolean circuits and supports boolean points~\cite{pullonen2015combining}, as such they are well suited for comparison, division and multiplication operations. As mentioned in~\cite{aliasgari2012secure} to date, the majority of techniques (PPML) are implemented using integers, while there are undeniable limitations to integer arithmetic. In contrast, as can be seen in~\autoref{tab:smpcimplementationcomaprison}, the papers that we have surveyed for SMPC mostly use boolean points. 
\section{TakeAways}
\label{sec:takeaways}
Following the results of our work, we have observed the following 
key takeaways, which 
may instigate further research:
\begin{itemize}[leftmargin=0.2cm]
	\item \textbf{High computational and communication costs.} The main reason HE has not been adopted for 
	standardized use 
	is related to the high cost associated with model training. Since the amount of noise and the size of the ciphertext 
	increase during 
	mathematical operations on encrypted values, the 
	training time required for an encrypted model increases almost exponentially with every processed
	batch. Consequently, 
	though 
	accuracy remains high enough to 
	compare with plaintext models, the training time (while using HE), is multiple times longer than the plaintext counterpart. 
	
	Other contributions rely on SMPC through distributed architectures, trading off the computational performance and communication overhead. SMPC offers a significantly better level of security at the expense of costly cryptographic operations, leading to computational and communication cost increases. 
	For example, in SMPC each party has 
	minimal computational costs. However  
	communication between parties is required and this 
	can 
	cause increased 
	communication costs.  
	Both HE and SMPC need to modify the model structure to match the corresponding PPML protocols.
	This affects accuracy and hinders efficiency with existing frameworks. One can see that HE and SMPC have their own pros and cons in terms of security, effectiveness, efficiency and scalability. We believe that 
	hybrid approaches to training and inference, such as the ones proposed by POSEIDON~\cite{sav2020poseidon} can enjoy the benefits of each component, thus provisioning the optimal trade-off between ML task performance and privacy-preserving overhead.
	

	\item \textbf{Lack of open-source implementations.} While currently there is a gap between theoretical advances and real-world applications, there are certain open-source projects and tools dedicated to PPML~\cite{henecka2010tasty, ryffel2018generic,dahl2018private}.  
OSI is the foundation of 
a lot of the modern scientific research -- providing reliability, flexibility, transparency, and opportunities for collaboration.
However, as 
	can be seen in 
	\autoref{table:results-of-he}, OSI of HE protocols are few and far between as most implementations are private for various copyright or personal reasons. Also, as shown in~\autoref{tab:smpcimplementationcomaprison}, 8 of the 15 works provide OSIs for SMPC, while the remaining 
do not,
but as stated in~\autoref{subsubsec:smpcprotocols}, there are some other issues relating to 
	protocol implementation, such as lack of proper documentation. Because of the sheer difficulty of the implementation task, most researchers only cover the results documented in the original paper, instead of recreating the implementations. 
	This in turn causes problems, such as 
	inability to reproduce the results of the papers and test the implementation in other environments or on different datasets. 
    One of the main contributions of this work is to highlight the importance of OSI in PPML 
    and the need for reproducibility and usability considerations in research. We aim to bridge the gap by encouraging researchers and practitioners to prioritize open-source practices and share their implementations as this would provide:
 \begin{itemize}[leftmargin=0.2cm]
     \item \underline{Reproducibility and Progress}: OSI provide opportunity to other researchers to reproduce and validate the results 
     of a paper. By sharing the code, researchers contribute to the transparency and integrity of the scientific process, allowing others to verify and build upon their work. This leads to increased confidence and fosters a culture of reproducibility. Additionally, OSI facilitate collaboration and knowledge sharing. 
     When codes are made public, other researchers can build upon the existing work, enhance it, and develop new techniques more rapidly. This accelerates the progress in the field of PPML by leveraging the collective efforts and expertise of the research community. 
     \item \underline{Long-term sustainability and wider impact}: OSI are often maintained and supported by a community of contributors, ensuring their long-term sustainability. By encouraging researchers, to share their code, it promotes the continuous development, improvement, and maintenance of PPML implementations, thus  
     addressing issues such as software bugs, compatibility with new platforms or libraries, and evolving security requirements, ensuring the longevity and usefulness of the implementations over time. Additionally, it lowers the barrier to entry for those interested in utilizing PPML for practical purposes.
 \end{itemize}
	
	\item \textbf{Lack of 
		possible attack analysis on implementations.} Every proposed implementation of HE 
	covered here assumes that the server is a semi-honest third-party and can only listen and watch the results of the training and inference without attempting to gain any additional information through malicious means. As mentioned previously, the assumption is realistic in most real-world scenarios 
	involving the use of cloud-based server providers. It does, however, 
	reveal a lack of analysis 
	when it comes to potential ML attacks 
	on the protocols. As a result, the lack of analysis on potential attacks 
	points to a possible gap in the security of HE, when 
	used in PPML. That is due to the fact that ML tasks open new vectors of attack for malicious actors. Similarly, most of the SMPC protocols covered in this study consider the honest, but curious model, however, some studies also take the malicious model into account. While the works we investigated assume a non-colluding server, collusion between the parties must be taken into account in 
    all of these adversary models. In SMPC, collusion is unavoidable and poses a severe privacy concern, because it allows parties to learn each other's sensitive private input. Typically, parties in collusion share their data and function parameter settings with one another. Therefore, it is crucial to weight privacy against collusion, while creating an effective SMPC protocol.
	
	\item \textbf{Lack of privacy-conscious regulatory frameworks} 
The available privacy-related regulations may require companies to announce that they are collecting 
all data and 
possibly provide users with the choice to opt out of said data gathering. However this 
appears to be a zero-sum game. 
Privacy policies can help data owners 
determine which data is shared, under what conditions, with whom, and for what purposes. It is essential to ascertain whether the policy will be implemented on the client or server side, 
which would change it into a format, where the data owner could allow or restrict access to other users and the server for certain purposes (such as marketing) or at least remove access 
users 
posing a possible privacy threat.
\end{itemize}
\section{Challenges and Future Directions}
\label{sec:challenges} 
Despite the aforementioned techniques for protecting private data, while performing ML training, non-privacy ML algorithms are still frequently employed, and private data is still transferred to the cloud. To date, there is no silver-bullet technique, when it comes to achieving privacy in ML. The privacy degree offered by the techniques we presented varies a lot depending on many factors such as the ML algorithm used, the adversary's capabilities and resources etc. Below we will discuss in detail the challenges of existing PPML techniques and their possible solutions. 

While ML advancements are frequently offered, the PPML techniques covered in this article are linked to specific ML algorithms. Therefore, PPML techniques are required to cope with the most recent advancements in ML. This creates a challenge, where both HE and SMPC need to modify the model structure to match the corresponding PPML protocols. This affects the training and inference accuracy and hinders compatibility with existing ML frameworks.

HE and SMPC allow work on encrypted data, thereby preserving the utility of the original datasets. However, 
their domain is quite limited, and scalability can be a major issue 
due to high costs. Both methods have differing advantages and disadvantages and as can be seen in the covered papers, contributions, 
relying on SMPC through distributed architectures, trade off lower computational costs with a higher communication overhead. SMPC offers a significantly better level of security at the expense of costly networking operations, for example, in SMPC each party has less computational cost 
compared to HE but requires a lot of communication between parties. This 
can, lead to high communication costs. On the other hand, in HE, the server 
incurs a substantial 
computational cost, as it has to train the entire model on its own, at almost no communication cost, as all of the data needed for training is sent to the server once. A possible way to combat the high costs of both techniques and achieve a higher degree of privacy may require the combination of multiple PPML techniques. Recent literature has proposed combinations of multiple PPML techniques, such as FL with HE~\cite{sav2020poseidon, choquette2021capc}, or DP with HE~\cite{jia2022blockchain, owusu2021msdp} or SMPC~\cite{mugunthan2019smpai, yuan2021label}, aiming for higher privacy guarantees and lower costs. 

This work also addresses the scarcity of OSI in PPML and makes concerted efforts to understand and overcome the underlying challenges. We investigate the reason behind this scarcity, including concerns related to security, compatibility, maintenance, and support. By raising awareness and emphasizing the benefits of OSI, we aim to encourage researchers to prioritize the sharing of PPML implementation. Additionally, we explore potential solutions and discuss the importance of collaboration, community building, and knowledge sharing to foster a culture of OSI in PPML.

In light of this, we conclude that these cryptographic approaches are still in a developmental phase and that, in the 
years to come, they will have developed 
enough to become an essential component of ML and cryptography. We conclude by summarizing our thoughts on potential future research directions 
involving both ML and PPTs that leverage as well as benefit multiple research communities such as ML, security, and privacy.

\begin{itemize}[leftmargin=0.2cm]
\item As covered in the sections 
above, performing computation on encrypted data using HE is computationally expensive. More specifically, training ML models requires a lot of mathematical operations and performing them 
on encrypted data, when training ML models, raises the computation costs to an impractical level. One possible solution is to use a hybrid approach -- HE and FE. The idea is to first encrypt the message using HE and then re-encrypt using symmetric FE. FE is used instead of 
HE for inner products and sum. This 
is comparatively faster and more efficient. After training, 
the message is first decrypted, following the same steps, with the FE symmetric key. It acquires the form of a HE ciphertext, and is 
further decrypted 
by use of the HE secret key with the aim of recovering the original plaintext.
\item DNN have millions of parameters. This 
is the main reason, why they are 
computationally expensive. 
Additionally, training a DNN on HE data 
increases the computing cost even more, sometimes to an unacceptable level. One possible approach for PPML is to use split learning (SL)~\cite{gupta2018distributed}, as it divides a DNN model so that part of the model is trained on the client side using plaintext data, and the remaining part 
is trained on the server side using encrypted data~\cite{khan2023split}. As part of the model is trained on plaintext, 
the DNN model's overall computation costs are lowered. Also, the privacy of user data is preserved as \begin{inparaenum}[\itshape a)\upshape] \item SL itself is believed to be a promising approach for 
	raw data protection, and \item 
	the computation on the server side is performed on top of encrypted data, hence not revealing 
    information about 
    client data.\end{inparaenum}
\item FL and SL have offered solutions to the privacy problems in ML, however, they are not complete and present various 
security problems and privacy leakages~\cite{ppflSurvey22, abuadbba2020can, li2021label}. 
HE can be used to solve the privacy leakage in SL and FL. Currently, there are two popular HE schemes that can be employed in ML, namely CKKS and TFHE, each with 
advantages and disadvantages. In the future, we would like to research the applications of these HE schemes in FL and SL and how they can be used in combination with each other as well as other privacy technologies in different scenarios. 
\item 
A great deal of work has recently been put into developing a reliable and secure protocol for ML tasks including SMPC. In SMPC, the parties constantly communicate with one another to jointly compute a function, 
thus adding overhead. An alternative to SMPC is to employ a hybrid strategy that combines SL and Function Secret Sharing (FSS)~\cite{boyle2016function}. The model's initial layers are trained using SL on the client side, while its remaining layers are trained using FSS on the server side. 
Sending the data to the layers is all that is required to execute the layers on the client side. However, FSS is used to construct the secret shares, for each layer or function on the server. 
\end{itemize}

\section{Conclusion}
\label{sec:conclusion}
As PPML has recently gained attention from both industry and academic researchers, in this work we provide a thorough analysis of SotA PPML based on HE and SMPC. First, we 
provide a general overview of the techniques used to implement privacy-preserving computation techniques on ML. We also describe various properties and settings pertaining to the privacy of ML models and data. This makes it possible to comprehend the 
entire range of PPML as well as the benefits and constraints of 
various sub-areas depending on the goals and settings. Second, we summarize the current SotA of PPML techniques and provide an analysis based on factors including the privacy goal, architecture, efficiency, and usability. 
With this method, we extracted recurring insights and flaws from the existing solutions. Third, we benchmarked and compared different PPML techniques, concerning privacy goals, communication, accuracy, and runtime to assess the viability of various PPML approaches for use in practical applications. Fourth, we list the key takeaways from our work, emphasizing the lessons learned and areas in need of further research effort. As such, this will help future researchers and practitioners apply PPML more effectively and have a better understanding of current limitations. Finally, we conclude that there are still many hurdles to overcome for HE and SMPC-based PPML to become practical. We outline the current challenges, that need to be addressed by the community, and discuss potential directions and approaches.


\begin{acks}
	This work was funded by the HARPOCRATES EU research project (No. 101069535).
\end{acks}

\bibliographystyle{ACM-Reference-Format}
\bibliography{sample-base}

\appendix

 \begin{table*}[t]
	\centering
	\caption{Comparison among privacy-preserving training methods using HE. All of the listed works assume a semi-honest threat model and do not take into account a potentially malicious server or client.}
        \label{table: hecomparison}
	\resizebox{0.95\textwidth}{!}{%
		\begin{tabular}{ccccccccccccccccccccclcccc} 
			\hline
			\multicolumn{2}{l}{\multirow{3}{*}{~ Framework}} & \multirow{3}{*}{Year} & \begin{turn}{75}Convolutional \end{turn} & \begin{turn}{75} Non-linear \end{turn} & \begin{turn}{75} Pooling \end{turn} &  & \begin{turn}{75} CKKS \end{turn} & \begin{turn}{75} FV \end{turn} & \begin{turn}{75} TFHE \end{turn} & \begin{turn}{75} BGV \end{turn} &  & \begin{turn}{75} SHE \end{turn} & \begin{turn}{75} LHE \end{turn} & \begin{turn}{75} FHE \end{turn} & \begin{turn}{75} MHE \end{turn}& &  & \begin{turn}{75} MNIST \end{turn} & \begin{turn}{75} CIFAR \end{turn} & \begin{turn}{75}UCI\end{turn} & \begin{turn}{75} iDASH \end{turn}\\ 
			\cline{4-6}\cline{8-11}\cline{13-16}\cline{19-22}
			\multicolumn{3}{c}{} & \multicolumn{3}{c}{Supported~ Layers} &  & \multicolumn{4}{c}{Techniques Used} & \multicolumn{7}{c}{HE Setting} & \multicolumn{4}{c}{Evaluation Dataset}\\
			\cline{4-16}\cline{19-22}
			\multicolumn{3}{c}{} & \multicolumn{15}{c}{Theoretical Metrics} & \multicolumn{6}{c}{Evaluation Metrics}\\
			\hline
			\multirow{6}{*}{HE} & Bonte \& Vercauteren & 2018 & \xmark & \cmark & \xmark &   & \xmark & \cmark & \xmark & \xmark &  & \cmark & \xmark & \xmark & \xmark & & & \xmark & \xmark & \xmark & \cmark\\
			& CryptoDL & 2019 &  \cmark & \cmark & \cmark &   & \xmark & \xmark & \xmark & \cmark &  & \cmark & \xmark & \xmark & \xmark & & & \cmark & \cmark & \cmark & \xmark \\
			& Nandakumar \textit{et al.} & 2019 &  \xmark & \cmark  & \xmark &   & \xmark & \xmark & \xmark & \cmark &  & \xmark & \xmark & \cmark & \xmark & & & \cmark & \xmark & \xmark & \xmark\\
			& PrivFT & 2019 &  \xmark & \xmark & \xmark &   & \cmark & \xmark & \xmark & \xmark &  & \xmark & \cmark & \xmark & \xmark & & & \xmark & \xmark & \cmark & \xmark\\
			& Mihara \textit{et al.} & 2020 &  \xmark & \cmark & \xmark &   & \cmark & \xmark & \xmark & \xmark &  & \xmark & \cmark & \xmark & \xmark & & & \xmark & \xmark & \cmark & \xmark\\
            & Sphinx & 2022 &  \cmark & \cmark & \cmark &   & \cmark & \xmark & \xmark & \xmark &  & \xmark & \cmark & \xmark & \xmark & & & \cmark & \cmark & \xmark & \xmark\\
			\hline
			\multirow{2}{*}{Hybrid} & Glyph & 2020 &  \cmark & \cmark & \cmark &   & \xmark & \xmark & \cmark & \cmark &  & \xmark & \cmark & \xmark & \xmark & & & \cmark & \xmark & \xmark & \xmark\\
			& POSEIDON & 2021 &  \cmark & \cmark & \cmark &  & \cmark & \xmark & \xmark & \xmark &  & \xmark & \xmark & \xmark & \cmark & & & \cmark & \cmark & \xmark & \xmark\\
			\hline
		\end{tabular}
	}
\end{table*}

\begin{table*}[t!]
	\centering
	\caption{Comparison 
		between privacy-preserving training methods using SMPC}
	\label{table:smpc comaprison}
	\resizebox{0.95\textwidth}{!}{%
		\arrayrulecolor{black}
		\begin{tabular}{ccccccccccccccclccccclccccccc} 
			\hline 
			\multicolumn{2}{l}{\multirow{3}{*}{\begin{tabular}[c]{@{}l@{}}\\~ Framework\end{tabular}}} & \multirow{3}{*}{Year} & \begin{turn}{75}Semi-honest\end{turn} & \begin{turn}{75}Malicious\end{turn} &  & \begin{turn}{75}Convolutional\end{turn} & \begin{turn}{75}ReLU \end{turn} & \begin{turn}{75}Maxpool \end{turn} & \begin{turn}{75} Linear \end{turn} &  & \begin{turn}{75} OT \end{turn} & \begin{turn}{75} GC \end{turn} & \begin{turn}{75} Secret Sharing\end{turn} & \begin{turn}{75} FSS \end{turn} & \multirow{3}{*}{} & \begin{turn}{75}LAN\end{turn} & \begin{turn}{75}WAN\end{turn} &  & \begin{turn}{75}MNIST\end{turn} & \begin{turn}{75}CIFAR\end{turn} & \begin{turn}{75}Tiny ImageNet\end{turn} & \begin{turn}{75}ImageNet\end{turn} &  & \begin{turn}{75}From SecureML\end{turn} & \begin{turn}{75}From MiniONN\end{turn} & \begin{turn}{75}LeNet\end{turn} & \begin{turn}{75}AlexNet\end{turn} & \begin{turn}{75}VGG16\end{turn} \\ 
			\cline{4-5}\cline{7-10}\cline{12-15}\cline{17-18}\arrayrulecolor{black}\cline{19-19}\arrayrulecolor{black}\cline{20-23}\cline{25-29}
			\multicolumn{2}{l}{} &  & \multicolumn{2}{c}{Threat Model} &  & \multicolumn{4}{c}{Supported~ Layers} &  & \multicolumn{4}{c}{Techniques Used} &  & \multicolumn{2}{c}{LAN/WAN} &  & \multicolumn{4}{c}{Evaluation Dataset} &  & \multicolumn{5}{c}{Neural~ Network~ Architectures} \\
			\cline{4-15}\cline{17-29}
			\multicolumn{2}{l}{} &  & \multicolumn{12}{c}{Theoretical Metrics} &  & \multicolumn{13}{c}{Evaluation Metrics} \\
			\hline
			
			\multicolumn{1}{c}{\multirow{4}{*}{2PC}} & SecureML & 2017 & \multicolumn{2}{c}{\halfcircL} &  & \cmark & \cmark & \cmark & \cmark &  & \cmark & \cmark  & \cmark & \xmark &  & \multicolumn{2}{c}{\fullcirc} &   & \cmark & \xmark & \xmark & \xmark &  & \cmark & \xmark & \xmark & \xmark & \xmark \\
			& Quotient & 2019 & \multicolumn{2}{c}{\halfcircL} &  & \cmark & \cmark & \cmark & \cmark &  & \cmark & \cmark & \cmark & \xmark &  & \multicolumn{2}{c}{\fullcirc} &  & \cmark & \xmark & \xmark & \xmark &  & \cmark & \xmark & \xmark & \xmark & \xmark \\
			& ParSecureML & 2020 & \multicolumn{2}{c}{\halfcircL} &  & \cmark & \cmark & \textendash & \cmark &  & \textendash & \textendash & \textendash & \xmark &  & \multicolumn{2}{c}{\textendash} &  & \cmark & \xmark  & \xmark & \xmark &  & \cmark & \xmark & \xmark & \xmark & \xmark \\ 
			& ABY2 & 2021 & \multicolumn{2}{c}{\halfcircL} &  & \cmark & \cmark & \cmark & \cmark &  & \cmark & \cmark & \cmark & \xmark &  & \multicolumn{2}{c}{\fullcirc} &  & \cmark & \xmark & \xmark & \xmark &  & \cmark & \xmark & \xmark & \xmark & \xmark  \\
           & AriaNN & 2022 & \multicolumn{2}{c}{\halfcircL} &  & \cmark & \cmark & \cmark & \cmark &  & \textendash & \cmark & \textendash & \cmark &  & \multicolumn{2}{c}{\fullcirc} &  & \cmark & \cmark & \xmark & \cmark &  & \xmark & \xmark & \cmark & \cmark & \cmark  \\
			\hline
			\multirow{7}{*}{3PC} & ABY\textsuperscript{3} & 2018 & \multicolumn{2}{c}{\fullcirc} &  & \cmark & \cmark & \cmark & \cmark &  & \cmark & \cmark & \cmark & \xmark &  & \multicolumn{2}{c}{\fullcirc} &  & \cmark & \xmark & \xmark & \xmark &  & \cmark & \cmark & \xmark & \xmark & \xmark \\

           & SecureNN & 2019 & \multicolumn{2}{c}{\fullcirc} &  & \cmark & \cmark & \cmark & \cmark &  & \xmark & \xmark & \cmark & \xmark &  & \multicolumn{2}{c}{\fullcirc} &  & \cmark & \xmark & \xmark & \xmark &  & \cmark & \cmark  & \cmark & \xmark & \xmark \\
			& BLAZE & 2020 & \multicolumn{2}{c}{\fullcirc} &  & \cmark & \cmark & \cmark & \cmark &  & \cmark & \cmark & \cmark & \xmark &  & \multicolumn{2}{c}{\fullcirc} &  & \xmark & \xmark & \xmark & \xmark &  & \xmark & \xmark & \xmark & \xmark&\xmark\\
			& Falcon & 2020 & \multicolumn{2}{c}{\fullcirc} &  & \cmark & \cmark & \cmark & \cmark &  & \xmark & \xmark & \cmark & \xmark &  & \multicolumn{2}{c}{\fullcirc} &  & \cmark & \cmark & \cmark & \textendash  &  & \cmark & \cmark & \cmark & \cmark & \cmark \\
			& SWIFT & 2021 & \multicolumn{2}{c}{\fullcirc} &  & \cmark & \cmark & \cmark & \cmark &  & \xmark & \xmark & \cmark & \xmark &  & \multicolumn{2}{c}{\halfcircR} &  & \cmark & \cmark & \xmark & \xmark &  & \xmark & \xmark & \cmark & \xmark & \cmark  \\
			& Adam in Private & 2021 & \multicolumn{2}{c}{\fullcirc} &  & \cmark & \cmark & \cmark & \cmark &  & \xmark  & \xmark & \cmark & \xmark &  & \multicolumn{2}{c}{\fullcirc} &  & \cmark & \cmark & \xmark & \xmark  &  & \cmark & \xmark & \xmark & \cmark & \cmark \\ 
			& CryptGPU & 2021 & \multicolumn{2}{c}{\halfcircL} &  & \cmark & \cmark & \cmark & \cmark &  & \xmark & \xmark & \cmark & \xmark &  & \multicolumn{2}{c}{\halfcircL} &  & \cmark & \cmark & \cmark  & \cmark &  & \xmark & \xmark & \cmark & \cmark & \cmark \\
			& Piranha & 2022 & \multicolumn{2}{c}{\halfcircL} &  & \cmark & \cmark & \cmark & \cmark &  & \xmark & \xmark & \cmark & \xmark &  & \multicolumn{2}{c}{\fullcirc} &  & \cmark & \cmark & \xmark  & \xmark &  & \cmark & \xmark & \cmark & \cmark & \cmark \\
			\hline
			\multicolumn{1}{c}{\multirow{5}{*}{4PC}}  
			& Trident & 2020 & \multicolumn{2}{c}{\fullcirc} & & \cmark & \cmark & \cmark & \cmark &  & \cmark & \cmark & \cmark & \xmark  &  & \multicolumn{2}{c}{\fullcirc} &  & \cmark & \xmark & \xmark & \xmark &  & \cmark & \xmark & \xmark & \xmark & \xmark \\
			& Fantastic Four & 2021 & \multicolumn{2}{c}{\fullcirc} &  & \cmark & \cmark & \cmark & \cmark &  & \xmark & \xmark & \cmark & \xmark &  & \multicolumn{2}{c}{\halfcircR} &  & \cmark & \xmark & \xmark & \textendash &  & \xmark & \xmark & \xmark & \xmark & \xmark \\
			& MPCLeague & 2021 & \multicolumn{2}{c}{\fullcirc} &  & \cmark & \cmark & \cmark & \cmark &  & \textendash & \cmark & \cmark & \xmark &  & \multicolumn{2}{c}{\halfcircR} &  & \textendash & \textendash & \textendash & \textendash &  & \textendash & \textendash & \cmark &  & \cmark \\
			& Tetrad & 2021 & \multicolumn{2}{c}{\fullcirc} &  & \cmark & \cmark & \cmark  & \cmark &  & \cmark & \cmark & \cmark & \xmark &  & \multicolumn{2}{c}{\halfcircR} &  & \cmark & \cmark & \xmark & \xmark &  & \cmark & \xmark & \cmark & \xmark & \cmark  \\
			\hline
		\end{tabular}
	}
\end{table*}

\begin{table*}[t!]
	\centering
	\caption{Comparison of training time, accuracy and 
		details for privacy-preserving training methods using HE. * -- None of the listed papers provide open-source code for their implementation.}
	\label{table:results-of-he}
	\resizebox{0.95\textwidth}{!}{
		\begin{tabular}{cccccccccccccccclcccc} 
			\hline
			\multicolumn{2}{l}{\multirow{0}{*}{~ Framework}} & \multirow{0}{*}{Year} &  & \begin{turn}{75}Training time(h) \end{turn} & \begin{turn}{75}Training accuracy(\%)\end{turn} &    & \begin{turn}{75} CPU \end{turn} & \begin{turn}{75} GPU \end{turn} & \begin{turn}{75} RAM(GB) \end{turn} & \begin{turn}{75} Environment \end{turn} & & & \begin{turn}{75}Algorithms\end{turn}\\
			\cline{5-6}\cline{8-11}\cline{13-16}
			\multicolumn{4}{c}{} & \multicolumn{2}{c}{Performance} & \multicolumn{6}{c}{Specifications} & \multicolumn{4}{c}{Availability*}\\
			\hline
			\multirow{6}{*}{HE} & Bonte \& Vercauteren & 2018 &  & 0.367-0.75 & 62.98-94.16 &    & - & - & - & - & & & \cmark \\
			& CryptoDL & 2019 &  & 2.91 & 99.0  &  & 12-core CPU & - & 48 & Ubuntu 14.04 (VM) & & &  \cmark\\
			& Nandakumar \textit{et al.} & 2019 &  & 60K-846K & 96.4-97.8 &  & Intel Xeon E5-2698 v3 & - & 250 & Linux & & & \xmark\\
			& PrivFT & 2019 &  & 120.96 & 91.49-98.80 &  & Intel Xeon E5-2620 & DGX-1 V100 & 180 & ArchLinux (CPU) & & &  \cmark\\
			& Mihara \textit{et al.} & 2020 &  & 29.8 & 98.47 &  & Intel Core i7-8700K & - & 32 & - &  & & \cmark\\
                & Sphinx & 2022 &  & 1560 & 72.0-96.0 &  & Intel Xeon E5-2683 v4 & - & 128 & Ubuntu 18.04.5 LTS &  & & \cmark\\
			\hline
			\multirow{2}{*}{Hybrid} & Glyph & 2020 &  & 192 & 98.6 &  & Intel Xeon E7-8890 v4 & - & 256 & - & & & \xmark\\
			& POSEIDON & 2021 &  & 1.47 & 80.2-96.9  &  & Intel Xeon E5-2680 v3 & - & 256 & Linux & & & \cmark\\
			\hline
		\end{tabular}
	}
\end{table*}

\begin{table*}[t!]
    \scriptsize
	\centering
	\caption{Comparison of training time and communication overhead, 
		while training various frameworks over popular benchmarking network architectures in the PPML domain on the MNIST dataset.  
		All networks and frameworks are exposed to a 15 epoch-training. Network 1 is a 3-layered fully-connected network from SecureML~\cite{mohassel2017secureml}, Network 2 is a 4-layered network from MiniONN~\cite{liu2017oblivious} that contains 2 convolution layers and 2 fully connected layers. Network 3 is LeNet~\cite{lecun1998gradient} with 2 convolution layers and 2 fully connected layers. }
	\label{tab:trainingResultsSMPCFrameworks}
	\resizebox{0.95\textwidth}{!}{
		\begin{tabular}{c|c|lll|lll|lll} 
			\hline
			\multirow{3}{*}{Number of parties} & \multirow{3}{*}{Framework} & \multicolumn{3}{c|}{Network 1 (SecureML)}                                 & \multicolumn{3}{c|}{Network 2 (MiniONN)}                                 & \multicolumn{3}{c}{LeNet}                                       \\ 
			\cline{3-11}
			&                            & \multicolumn{3}{c|}{Local host}                                & \multicolumn{3}{c|}{Local host}                                & \multicolumn{3}{c}{Local host}                                  \\ 
			\cline{3-11}
			&                            & \multicolumn{1}{c}{Time (sec)} & \multicolumn{1}{c}{Comm (MB)} & Accuracy & \multicolumn{1}{c}{Time (sec)} & \multicolumn{1}{c}{Comm (MB)} & Accuracy & \multicolumn{1}{c}{Time (sec)} & \multicolumn{1}{c}{Comm (MB)} & Accuracy  \\ 
			\hline
			2PC                                & SecureML                   &  \ \ \ \ \textendash      & \ \ \ \ \textendash              &      92.98    &  \ \ \ \ \textendash     &  \ \ \ \ \textendash      &  \ \ \ \ \textendash   & \ \ \ \ \textendash     &     \ \ \ \ \textendash  & \ \ \ \textendash \\  
			3PC                                & SecureNN                   &           \ \ \ \ \textendash               &            \ \ \ \ \textendash              &     \ \ \ \ \textendash     &         \textendash                 &          \textendash               &       \ \ \ \ \textendash   &        136.553      &    \ \ \  21058.3                    &     \ \ \ \textendash       \\
			3PC                                & Falcon                     & 2.38            & \ \ \ 256.57            &  \ \ \ \  \textendash     & 67.75              & \ 6944.4              & \ \ \ \ \textendash        & 173.9             & \ \ \ 1800                &     \ \ \ \textendash      \\
			2PC                                & AriaNN                     &      5.38 (h)                   &       38757.89                   &    97.96  &      12.81 (h)                   &       868308.48                  &     97.92     &                14.15  (h)         &          1324597.248             &    89.36 \\ \hline 
		\end{tabular}
	}
\end{table*}

\begin{table*}[tp!]
\scriptsize
	\centering
	\caption{Comparison among privacy-preserving training methods using SMPC}
	\label{tab:smpcimplementationcomaprison}
		\resizebox{0.95\textwidth}{!}{
	\begin{tabular}{ccccccclccccclcccccc} 
		\hline
		\multicolumn{2}{l}{\multirow{3}{*}{\begin{tabular}[c]{@{}l@{}}\\~ Framework\end{tabular}}} & \begin{turn}{75}Input/Output\end{turn} & \begin{turn}{75}Model \end{turn} &  & \begin{turn}{75}GOD\end{turn} & \begin{turn}{75}Fairness\end{turn} & \begin{turn}{75}Abort\end{turn} &  & \begin{turn}{75}Integer\end{turn} & \begin{turn}{75}Boolean\end{turn}  & \begin{turn}{75}Floating Point\end{turn} & \multirow{3}{*}{} & \begin{turn}{75}Public\end{turn} & \begin{turn}{75}Private\end{turn} &  & \begin{turn}{75}Accuracy\end{turn} & \begin{turn}{75}Complexity\end{turn} & \begin{turn}{75}Communication 
		\end{turn} & \begin{turn}{75}Time\end{turn} \\ 
		\cline{3-4}\cline{6-8}\cline{10-12}\cline{14-15}\cline{17-20}
		\multicolumn{2}{l}{} & \multicolumn{2}{c}{Privacy} &  & \multicolumn{3}{c}{Security} &  & \multicolumn{3}{c}{Data types} &  & \multicolumn{2}{c}{Implementation} &  & \multicolumn{4}{c}{Parameters for Comparison} \\ 
		\cline{3-12}\cline{14-20}
		\multicolumn{2}{l}{} & \multicolumn{10}{c}{Theoretical Metrics} &  & \multicolumn{7}{c}{Implementation} \\ 
		\hline
		\multirow{4}{*}{2PC} & SecureML & \multicolumn{2}{c}{\halfcircL} &  & \textendash & \textendash & \textendash &  & \textendash & \cmark & \cmark &  & \multicolumn{2}{c}{\halfcircL} &  & \cmark & \cmark  & \cmark  & \cmark  \\
		& Quotient & \multicolumn{2}{c}{\halfcircL} &  & \textendash & \textendash & \textendash &  & \xmark & \cmark & \xmark &  & \multicolumn{2}{c}{\halfcircR} &  & \cmark & \cmark & \cmark & \cmark \\
		& ParSecureML & \multicolumn{2}{c}{\halfcircL} &  & \textendash & \textendash & \textendash &  & \textendash & \textendash & \textendash &  & \multicolumn{2}{c}{\halfcircL} &  & \textendash & \cmark & \cmark & \cmark \\ 
  		& ABY\textsuperscript{2} & \multicolumn{2}{c}{\halfcircL} &  & \textendash & \textendash & \textendash &  & \cmark & \cmark & \textendash &  & \multicolumn{2}{c}{\halfcircR} &  & \textendash & \cmark & \cmark & \cmark \\ 
		& AriaNN & \multicolumn{2}{c}{\fullcirc} &  & \textendash & \textendash & \textendash &  & \textendash & \cmark & \textendash &  & \multicolumn{2}{c}{\halfcircL} &  & \cmark & \cmark & \cmark & \cmark \\
		\hline
		\multirow{7}{*}{3PC} & SecureNN & \multicolumn{2}{c}{\halfcircL} &  & \xmark  & \xmark & \cmark &  & \xmark & \cmark & \xmark &  & \multicolumn{2}{c}{\halfcircL} &  & \cmark & \cmark & \cmark & \cmark \\
		& ABY\textsuperscript{3} & \multicolumn{2}{c}{\halfcircL} &  & \xmark & \xmark & \cmark &  & \xmark & \cmark & \xmark &  & \multicolumn{2}{c}{\halfcircL} &  & \cmark & \textendash & \cmark & \cmark \\
		& BLAZE & \multicolumn{2}{c}{\halfcircL} &  & \xmark & \cmark & \xmark &  & \cmark & \cmark & \textendash &  & \multicolumn{2}{c}{\halfcircR} &  & \xmark & \cmark & \cmark & \xmark \\
		& Falcon & \multicolumn{2}{c}{\halfcircL} &  & \xmark & \xmark & \cmark &  & \cmark & \textendash & \textendash &  & \multicolumn{2}{c}{\halfcircL} &  & \cmark & \cmark & \cmark & \cmark \\
		& SWIFT & \multicolumn{2}{c}{\halfcircL} &  & \cmark & \xmark & \xmark &  & \textendash & \cmark & \textendash  & & \multicolumn{2}{c}{\halfcircR} &  & \xmark & \cmark & \cmark & \cmark \\
		& Adam in Private & \multicolumn{2}{c}{\halfcircL} &  & \xmark & \xmark & \cmark &  & \cmark & \cmark & \cmark &  & \multicolumn{2}{c}{\halfcircR} &  & \cmark & \cmark & \cmark & \cmark \\ 
		& CryptGPU & \multicolumn{2}{c}{\halfcircL} &  & \textendash & \textendash & \textendash &  & \textendash & \xmark & \cmark &  & \multicolumn{2}{c}{\halfcircL} &  & \cmark & \cmark & \cmark & \cmark \\
		& Piranha & \multicolumn{2}{c}{\halfcircL} &  & \textendash & \textendash & \textendash &  & \textendash & \textendash & \textendash &  & \multicolumn{2}{c}{\halfcircL} &  & \cmark & \cmark & \cmark & \cmark \\
  
  \hline
		\multirow{4}{*}{4PC} & Trident & \multicolumn{2}{c}{\halfcircL} &  & \xmark & \cmark & \xmark &  & \xmark & \cmark & \textendash &  & \multicolumn{2}{c}{\halfcircR} &  & \xmark & \cmark & \cmark & \cmark \\
		& Fantastic Four & \multicolumn{2}{c}{} &  & \cmark & \xmark & \xmark &  & \cmark & \cmark & \textendash &  & \multicolumn{2}{c}{\halfcircL} &  & \cmark & \cmark & \cmark & \cmark \\
		& MPCLeague & \multicolumn{2}{c}{\halfcircL} &  & \cmark & \xmark & \xmark &  & \cmark & \cmark & \textendash &  & \multicolumn{2}{c}{\halfcircR} &  & \textendash & \cmark & \cmark & \cmark \\
		& Tetrad & \multicolumn{2}{c}{\halfcircL} &  & \cmark & \cmark & \xmark &  & \cmark & \cmark & \cmark &  & \multicolumn{2}{c}{\halfcircR} &  & \xmark & \cmark & \cmark & \cmark \\
		\hline
	\end{tabular}
}
\end{table*}
\section{Preliminaries}
\label{sec: preliminaries}
In this section, we discuss the categorization, construction, 
and schemes of HE. 
Additionally, we present the basic primitives developed for constructing SMPC protocols with different properties, security notations, and dimensions. We conclude this section by discussing the various ML models and datasets that PPML employs\footnote{Due to space constraints, ``Machine Learning and Datasets'' are in appendix section.}.

\subsection{Homomorphic Encryption}
\label{subsec: HE}


There are four main categories of HE based on the scheme's ability to perform various mathematical calculations. The main categories are: \textit{Partially Homomorphic Encryption (PHE)}~\cite{paillier1999public, cohen1985robust, shafi1982probabilistic, elgamal1985public, rivest1983method}, \textit{Fully Homomorphic Encryption (FHE)}~\cite{gentry2009fully, viand2021sok, erabelli2020pyfhe, chillotti2016faster, chillotti2020tfhe}, \textit{Somewhat Homomorphic Encryption (SHE)}~\cite{armknecht2015guide, dijk2010fully} and \textit{Leveled Homomorphic Encryption (LHE)}~\cite{armknecht2015guide, brakerski2014leveled, gentry2013homomorphic, cheon2017homomorphic}.
FHE schemes have the highest potential in various applications. However, their use and implementation is heavily stunted by the complex mathematics 
required to implement them~\cite{acar2018survey}. 
FHE's mathematics require a different approach to encryption and decryption compared to the ones provided in prior PHE schemes, such as RSA or Elgamal. 
The major breakthrough that was in order 
in the field of HE occurred when 
Gentry proved that ciphertext homomorphism can be achieved through the use of lattices~\cite{gentry2009fully}. The lattice-based cryptography 
in Gentry's work used 
the security provided in the Learning with Error (LWE) problem described by Oded Regev~\cite{regev2010learning}. Future schemes of HE continued to rely on the high security provided by the LWE assumption and its variations (DLWE~\cite{regev2009lattices}, ring LWE~\cite{lyubashevsky2010ideal}, etc). 
However, because of some major drawbacks in FHE, researchers, including Gentry himself, 
focused on alternative solutions. 
This 
led to the 
SHE and LHE schemes mentioned above, which, 
though based on the same concept, 
lowered the complexity of the algorithms by limiting the depth of the circuit through scheme parameters or a predefined value to avoid bootstrapping~\cite{cabrero2021sok,naehrig2011can,acar2018survey, brakerski2014leveled}.

Generally, a HE scheme can be defined as~\cite{gentry2009ideal}:
\begin{definition}[Homomorphic Encryption]
Let $\mathsf{HE}$ be a (public-key) homomorphic encryption scheme with a quadruple of PPT algorithms $\mathsf{HE = (KeyGen, } \allowbreak \mathsf{Enc, Dec, Eval)}$ such that:
\end{definition}
\begin{itemize}
\item $\mathbf{HE.KeyGen:}$ The key generation algorithm $\left(\mathsf{pk, evk, sk}\right) \leftarrow \mathsf{HE.KeyGen}(1^{\lambda})$ takes as input a unary representation of the security parameter $\lambda$, and outputs  a public key $\mathsf{pk}$, an evaluation key $\mathsf{evk}$ and a private key $\mathsf{sk}$.
\item $\mathbf{HE.Enc:}$ The encryption algorithm ${c \leftarrow \mathsf{HE.Enc}(\mathsf{pk}, x)}$ takes as input the public key $\mathsf{pk}$ and a message $x$ and outputs a ciphertext $c$.
\item $\mathbf{HE.Eval:}$ The evaluation algorithm $c_f \leftarrow \mathsf{HE.Eval}({\mathsf{evk}}, f, \allowbreak c_1, \dots, c_n)$ takes as input the evaluation key $\mathsf{evk}$, a function $f$, and a set of $n$ ciphertexts, and outputs a ciphertext $c_f$.
\item $\mathbf{HE.Dec:}$ The decryption algorithm $\mathsf{HE.Dec}({\mathsf{sk}}, c) \rightarrow x$, takes as input the secret key $\mathsf{sk}$ and a ciphertext $c$, and outputs a plaintext $x$.
\end{itemize}

\textit{\underline{Limitation}:} 
Despite advancements over time, all HE techniques still require high-end machines to properly implement bootstrapping and are significantly more computationally expensive when compared to other privacy-preserving alternatives. Furthermore, SHE and LHE schemes limit the number of operations performed on a ciphertext, which might result in lower accuracy and reduced functionality in certain applications.

\noindent \textbf{\textit{HE schemes:}} HE includes a variety of encryption schemes that can perform various types of computations over encrypted data. These encryption schemes define, which operations are available and the type of activation and architectures that can be used. The common encryption schemes BGV~\cite{brakerski2014leveled}, and FV~\cite{fan2012somewhat} (see~\autoref{fig:hetaxonomy}) share many similarities such as Single Instruction Multiple Data (SIMD) operations 
with an integer-only message space and support bootstrapping~\cite{viand2021sok}. The two other popular schemes, CKKS~\cite{cheon2017homomorphic} and TFHE~\cite{chillotti2016faster}, support bootstrapping (
which is faster in TFHE). The CKKS scheme uses real numbers as the message space and TFHE only supports single bits 
in the message space. However, the fact that TFHE does not support SIMD operations, makes it less attractive for 
matrix multiplication. Another major HE scheme, not used by any of the papers, is FHEW~\cite{ducas2015fhew}. FHEW is an FHE scheme, which allows homomorphic NAND operations on two encrypted bits $\mathsf{E(b_1)}$ and $\mathsf{E(b_2)}$ to get the resulting bit $\mathsf{E(b_1 \barwedge b_2)}$. 
As the scheme is primarily built for only NAND operations, it has limited uses in more complex applications. However, the main benefit of FHEW is that it allows much faster bootstrapping, when compared to other HE schemes.


\subsection{Secure Multi-party Computation}
\label{subsec: mpc}

To properly apply SMPC techniques to ML, it is important to manage the costs of the SMPC frameworks, as ML requires numerous rounds of communication between the participant and the server. This may lead to 
significant communication costs. As such, various techniques have been developed over the years to help construct cost-effective SMPC frameworks, such as:

%

\noindent \textbf{\textit{Secret Sharing (SS):}} 
Splitting a secret into multiple shares and distributing each share to a different party. The secret can only be recovered when all parties 
join shares~\cite{shamir1979share}. 

\noindent\textit{\underline{Limitation}:}  In SS, multiple parties use a protocol 
to jointly compute a function on their inputs. As SS requires communication between parties, it can lead to high communication cost.

%

\noindent \textbf{\textit{Oblivious Transfer (OT):}} 
OT is a protocol where a sender sends one of several possible pieces of information to a receiver without knowing which piece of information the receiver obtained. 

\noindent\textit{\underline{Limitation}:} In OT protocols, the amount of data transmitted is proportional to the bit length of the input data. This implies a considerable amount of communication complexity. 
The communication overhead can cause significant WAN delays. 

\noindent\textbf{\textit{Garbled Circuits (GC):}} 
In~1986, Andrew Yao proposed GC~\cite{Yao1982ProtocolsFS} -- a cryptographic protocol that enables two parties to jointly evaluate a function over their private inputs without the presence of a trusted party. Yao's GC transforms any function into a securely-evaluated function 
by modeling 
it as a Boolean circuit. The inputs and outputs of each gate are masked so that the party executing the function cannot discern any information about the inputs or intermediate values of the function. 

\noindent\textit{\underline{Limitation}:}  The computation and communication overhead of ML execution utilizing the GC protocol is controlled by the number of neurons in each ML layer~\cite{rouhani2018deepsecure}. Each ML model layer contains thousands of values, which may result in large computing and communication overheads. 

\noindent \textbf{\textit{Security Notions:}} 
In SMPC, a set of mutually distrusting parties wish to jointly and securely compute a function of their inputs. This computation should be 
performed in such a way that each party obtain the proper result, and none of the parties learn anything beyond their prescribed output. 
A precise definition of security is required in order to prove that the SMPC protocol provides secure computation. 
A number of definitions have been proposed to ensure several security parameters 
that encompass most SMPC tasks. Here, we discuss 
this set of security parameters~\cite{zhao2019secure, lindell2020secure}.

\begin{itemize}[leftmargin=0.2cm]
	\item \textbf{Privacy} -- No party should learn 
	more than its prescribed output. More specifically, only what can be inferred from the output itself should be known about other parties' inputs.
	\item \textbf{Correctness} -- Each party is guaranteed that the output 
	it receives is correct.
	\item \textbf{Independence of input} -- The corrupt parties must choose their inputs independently of the honest parties' inputs.
	\item \textbf{Robustness} -- Regardless of the adversary's actions, all parties can compute the protocol's output. There are various levels of robustness:
	\begin{itemize}[leftmargin=0.4cm]
		\item \textit{Guaranteed Output Delivery (GOD)} -- The strongest level is GOD, where honest parties are always certain to receive the output regardless of 
		the adversary's behaviors.
		\item \textit{Fairness} -- It is a weaker variant compared to GOD. It states that the adversary receives the output if and only if the honest parties receive the output.
		\item \textit{Security with Selective Abort} -- It is the weakest security notion, where the adversary can selectively deprive the honest parties from 
		the output. In selective abort, at the end of computation, the adversary 
		receives the output for certain values, but can prevent 
		honest parties from 
		receiving their outputs.
	\end{itemize}
\end{itemize}

Using these security concepts, we later discuss the privacy and security guarantees offered by the current PPML protocols.
\section{Machine Learning and Datasets}
\label{ssec: mlanddatasets}
ML automates the analysis of datasets, producing models that reflect general relationships as found in the data. The two fundamental phases of ML are the training 
where the model is trained on input data, and the inference stage, where the trained model is put to use. ML techniques are divided into three classes, characterized by the nature of data available for analysis~\cite{chatzilygeroudis2021machine}. 

\begin{itemize}[leftmargin=0.2cm]
	\item \textbf{Supervised learning} -- In this type, 
	each input training example has a corresponding output that is also referred to as 
	label. The objective is to train a model that can map the input examples to their outputs as accurately as possible. Examples of applications using supervised learning include: Image classification, text classification, spam filtering, machine translation, etc~\cite{singh2016review}. 
	\item \textbf{Unsupervised learning} -- 
	Uses ML algorithms to analyze and cluster unlabelled datasets. These algorithms discover hidden patterns in data without the need for human intervention. Examples of applications using unsupervised learning include: Recommendation system, anomaly detection, etc~\cite{khanum2015survey}.
	\item \textbf{Reinforcement learning} -- It is neither based on supervised 
	nor unsupervised learning. The algorithm learns by exploring 
	the environment and taking actions, 
	thus maximizing 
	cumulative rewards. 
	It works with data in 
	sequences of actions, observations, and rewards. Examples of reinforcement learning applications can be found in 
	a plethora of areas such as self-driving cars, gaming and healthcare~\cite{wiering2012reinforcement}. 
\end{itemize}

Following, we present a list of popular ML models that are trained in a privacy-preserving manner.


\begin{itemize}[leftmargin=0.2cm]
	\item \textbf{Linear Regression (LR)} --
	This ML approach 
	models the relationship between two variables: 
	a dependent variable $y$ and an independent variable $x$. If a model 
	only features 
	an independent variable, then 
	it is called 
	a simple LR model; if it features 
	more than one, then 
	it is called multi-linear regression~\cite{seber2012linear}. 
	\item \textbf{Convolutional Neural Network (CNN)} --
	It is a Deep Neural Network (DNN) 
	often used in matrix-based applications like image recognition. As 
	the name indicates, CNN uses convolution operations
	performed on the input data with 
	a filter or kernel to produce feature maps~\cite{albawi2017understanding}. 
	\item \textbf{Logistic Regression (LoR)} --
	A generalized regression model 
	modeling the probability of a discrete outcome given an input variable. It makes use of a logistic function and can be used to describe 
	certain nonlinear relations~\cite{kleinbaum2002logistic}. The most common LoR models 
	produce a binary outcome, that is a result that can assume 
	two values 
	(yes/no, true/false etc.). Multinomial LoR can be used to model scenarios with more than two discrete outcomes.
	\item \textbf{LeNet}~\cite{lecun2015lenet} -- It is a CNN architecture that was first proposed by LeCun \textit{et al.} and was used in the automatic detection of zip codes and digit recognition. The network contains~2 convolutional and~2 fully connected layers. 
 LeNet is the first architecture that shows the application of CNNs on a real-world dataset.
	
	\item \textbf{AlexNet}~\cite{krizhevsky2012imagenet} -- This network is the winner of the ImageNet ILSVRC-2012 competition~\cite{ImageNet:2012}. 
	It consists of~5 convolutional layers and~3 fully connected layers,  
	uses a batch normalization layer (for stability and efficient training), and has about~60 million parameters. 
	Achieved an impressive result of about 10.8\% lower than the runner up for a top-5 error. This 
	was accomplished by training the deep CNN using graphics processing units (GPUs). AlexNet made a big impact in computer vision and artificial intelligence, and the original paper has accumulated more than 110,000 citations according to google scholar at the time of writing this paper.
	
	\item \textbf{VGG16}~\cite{simonyan2014very} -- 
	Won the first place in the localisation task and second place in the classification task of the ILSVRC-2014 competition~\cite{ImageNet:2014}. 
	It consists of 16 layers and has about~130 million parameters. VGG16 shows 
	we can build very DNNs, e.g.\ 16-19 weight layers, by utilizing very small convolution filters (size $3\times 3$). VGG16 shows that very DNNs can generalize well to various datasets and reproduce SotA results.
\end{itemize}





Data is an essential component of the ML model. 
Below, we discuss popular datasets used for training the PPML techniques:

\begin{itemize}[leftmargin=0.2cm]
	\item \textbf{MNIST} --
	A collection of hand-written digits. Consists of~60,000 images in the training set and~10,000 in the test set. Each image is a~28~$\times$~28 pixel image, the element in the matrix ranges from zero to~255 along with a label between~0 and~9~\cite{lecun1998mnist}. 
	\item \textbf{CIFAR} --
	It consists of~60,000 
 colored images out of which~50,000 are used for training and the remaining~10,000 are used for testing. These images are grouped into~10 mutually exclusive classes: airplanes, automobile, bird, cat, deer, dog, frog, horse, ship, truck~\cite{krizhevsky2009learning}.
	\item \textbf{Tiny ImageNet} --
	It is a subset of the ImageNet dataset in the ILSVRC~\cite{ImageNet:Main}.
	Tiny ImageNet dataset consists of~100,000 training samples and~10,000 test samples with~200 different classes. Each sample is cropped to a size of 64~$\times$~64~$\times$~3~\cite{le2015tiny}. 
	\item \textbf{UCI} --
	The University of California, Irvine (UCI) ML repository~\cite{Dua:2019} is a collection of more than 600 datasets of varying sizes used in ML tasks to evaluate the effectiveness of an ML algorithm. Some of the more well-known datasets in this repository are the Iris dataset~\cite{fisher_1988} and YouTube Spam Collection~\cite{almeida_silva_santos_hidalgo_2017}.
	\item \textbf{iDASH} --
	iDASH Privacy \& Security Workshop holds an annual secure genome analysis competition, which uses a customized genetic dataset every year 
	to create PPML algorithms 
	that classify sensitive genome data~\cite{tang_wang_wang_jiang_2018}. 
	\item \textbf{Yelp Dataset} -- It contains 6,990,280 reviews of 150,346 businesses from the website Yelp, which publishes crowd-sourced reviews of various establishments~\cite{inc._2018}. 
	\item \textbf{AG News Classification Dataset} -- The dataset is constructed from 4 of the largest classes in \textit{AG's corpus of news articles}~\cite{gulli_ferragina_2005}. The original corpus contains over a million news articles from more than 2,000 news sources and was made by the prior academic news search engine \textit{ComeToMyHead}. The classification dataset contains 120,000 training samples and 7,600 testing samples~\cite{zhang_2020}. 
	\item \textbf{IMDB Dataset of Movie Reviews} -- Also known as the \textit{Large Movie Review Dataset}
	~\cite{maas-EtAl:2011:ACL-HLT2011,LargeMovie:Dataset} is a dataset for binary sentiment classification containing an equal amount of movie review samples for training and testing (50,000 samples).
	\item \textbf{DBPedia Ontology Classification Dataset} -- 
	A crowd-sourced collection of the most commonly used infoboxes within Wikipedia~\cite{community_2021}. 
	Dataset consists of 14 ontology classes and each class has 40,000 training samples and 5,000 testing samples~\cite{zhang_2015}. 
	As such, the total size of the training dataset is 560,000, plus 70,000 testing~\cite{zhang2015character}.
\end{itemize}

\end{document}